%% file: main.tex
\documentclass[acmsmall, natbib=false]{acmart}
\AtBeginDocument{%
  }

\RequirePackage[
  datamodel=acmdatamodel,
  style=acmnumeric,
  ]{biblatex}

\usepackage{amsmath}
\usepackage{amsfonts}
\usepackage{mathtools} % Enhances amsmath
\usepackage{booktabs} % High-quality tables
\usepackage{multirow} % Table cells spanning multiple rows
\usepackage{graphicx} % Graphics inclusion
\usepackage{subfig} % Subfigure support
\usepackage{subcaption} % Subfigure support
\usepackage[table, xcdraw]{xcolor} % Text and background colors
\usepackage{enumitem} % Customizable lists (itemize, enumerate)

\usepackage{array} % Additional table options
\usepackage{tabularx} % Flexible-width tables
\usepackage{float} % Improved float handling
\usepackage{adjustbox} % Align/resize elements (e.g., tables or figures)
\usepackage{ragged2e} % Justify and ragged text
\usepackage{wrapfig} % Wrapping text around figures
\usepackage{longtable} % Tables spanning multiple pages
\usepackage{ltablex} % Combines longtable and tabularx
\usepackage{siunitx} % Consistent number and unit formatting

\usepackage{soul} % Highlighting text
\usepackage[normalem]{ulem} % Underlining and strikeout

\usepackage{stfloats} % Enables positioning of floats (e.g., double-column figures)

\usepackage{tikz} % Drawing and graphics
    
\newcommand*\circled[1]{\tikz[baseline=(char.base)]{ \node[shape=circle,fill,inner sep=0.4pt] (char) {\textcolor{white}{#1}};}}

\usepackage{textcomp} % Additional text symbols
\usepackage{url} % URL formatting
\usepackage{subcaption}

\usepackage[strings]{underscore}

\usepackage[usenames,dvipsnames]{xcolor}
\usepackage[breakable, theorems, skins]{tcolorbox}
\tcbset{enhanced}
\DeclareRobustCommand{\mybox}[2][gray!10]{%
\begin{tcolorbox}[   %% Adjust the following parameters at will.
        breakable,
        left=0pt,
        right=0pt,
        top=0pt,
        bottom=0pt,
        colback=#1,
        colframe=black!70,
        width=\dimexpr\columnwidth\relax, 
        enlarge left by=0mm,
        boxsep=2pt,
        arc=0pt,outer arc=0pt,
        ]
        #2
\end{tcolorbox}
}

\newcommand{\AF}[1]{\textcolor{blue}{AF - #1 - AF}}

\begin{document}

\title{SynergAI: Edge-to-Cloud Synergy for Architecture-Driven High-Performance Orchestration for AI Inference}

\author{Foteini Stathopoulou}
\email{fstathopoulou@microlab.ntua.gr}
\orcid{0009-0008-9695-8905}
\affiliation{
  \institution{National Technical University of Athens}
  \streetaddress{Heroon Polytechniou 9}
  \city{Zografou}
  \country{Greece}
  \postcode{15780}
}

\author{Aggelos Ferikoglou}
\email{aferikoglou@microlab.ntua.gr}
\orcid{0009-0004-9188-2930}
\affiliation{
  \institution{National Technical University of Athens}
  \streetaddress{Heroon Polytechniou 9}
  \city{Zografou}
  \country{Greece}
  \postcode{15780}
}

\author{Manolis Katsaragakis}
\email{mkatsaragakis@microlab.ntua.gr}
\orcid{0000-0001-8116-3503}
\affiliation{
  \institution{National Technical University of Athens}
  \streetaddress{Heroon Polytechniou 9}
  \city{Zografou}
  \country{Greece}
  \postcode{15780}
}

\author{Dimosthenis Masouros}
\email{demo.masouros@microlab.ntua.gr}
\orcid{0000-0001-6147-6908}
\affiliation{
  \institution{National Technical University of Athens}
  \streetaddress{Heroon Polytechniou 9}
  \city{Zografou}
  \country{Greece}
  \postcode{15780}
}

\author{Sotirios Xydis}
\email{sxydis@microlab.ntua.gr}
\orcid{0000-0003-3151-2730}
\affiliation{
  \institution{National Technical University of Athens}
  \streetaddress{Heroon Polytechniou 9}
  \city{Zografou}
  \country{Greece}
  \postcode{15780}
}

\author{Dimitrios Soudris}
\email{dsoudris@microlab.ntua.gr}
\orcid{0000-0002-6930-6847}
\affiliation{
  \institution{National Technical University of Athens}
  \streetaddress{Heroon Polytechniou 9}
  \city{Zografou}
  \country{Greece}
  \postcode{15780}
}

\renewcommand{\shortauthors}{Stathopoulou et al.}

\begin{abstract}
The rapid evolution of Artificial Intelligence (AI) and Machine Learning (ML) has significantly heightened computational demands, particularly for inference-serving workloads.
While traditional cloud-based deployments offer scalability, they face challenges such as network congestion, high energy consumption, and privacy concerns.
In contrast, edge computing provides low-latency and sustainable alternatives but is constrained by limited computational resources.
In this work, we introduce \texttt{SynergAI}, a novel framework designed for performance- and architecture-aware inference serving across heterogeneous edge-to-cloud infrastructures.
Built upon a comprehensive performance characterization of modern inference engines, \texttt{SynergAI} integrates a combination of offline and online decision-making policies to deliver intelligent, lightweight, and architecture-aware scheduling.
By dynamically allocating workloads across diverse hardware architectures, it effectively minimizes Quality of Service (QoS) violations.
We implement \texttt{SynergAI} within a Kubernetes-based ecosystem and evaluate its efficiency.
Our results demonstrate that architecture-driven inference serving enables optimized and architecture-aware deployments on emerging hardware platforms, achieving an average reduction of $2.4\times$ in QoS violations compared to a State-of-the-Art (SotA) solution.

\end{abstract}






\maketitle

%%%%%%%%%%%%%%%%%%%%%%%%%%%%%%%%%%%%%%%%%%%%%%%%
\input{sections/00_Introduction}
\input{sections/01_Related_Work}

\input{sections/02_Characterization}

\input{sections/03_Scheduler}
\input{sections/04_Experimental}
\input{sections/05_Conclusion}

%%%%%%%%%%%%%%%%%%%%%%%%%%%%%%%%%%%%%%%%%%%%%%%%

%\bibliographystyle{ACM-Reference-Format}
%\bibliography{sample-base}
%\bibliographystyle{acmnumeric}
%\bibliography{bibliography}
\printbibliography

\end{document}

%% file: sections/00_Introduction.tex
\section{Introduction}
\label{sec:introduction}

In recent years, the rapid advancement of AI and ML applications and their widespread integration into daily life has brought the new era of intelligent computing.
Technologies such as Deep Learning (DL)~\cite{pouyanfar2018survey} and Dynamic ML~\cite{alali2022proficient} are being widely used for inference across diverse domains, including healthcare~\cite{shaheen2021applications}, computer vision~\cite{khan2021machine}, and finance~\cite{gogas2021machine}.  
Modern ML inference is characterized by the convergence of increasingly complex models, rapid adaptation to specialized domains, and growing end-user demands.
This combination results in significantly higher computing, memory, and storage requirements. 
For example, widely used SotA DL models like YOLOv8m~\cite{yolov8_ultralytics} have over 25 million parameters. 
Newer architectures, such as ViT-G/14~\cite{Radford2021LearningTV}, are even more demanding, reaching nearly 2 billion parameters.
As a result, these models place heavy demands on system resources and can lead to significant performance bottlenecks during deployment.

Traditionally, these models are governed by throughput- and latency-oriented QoS~\cite{ferikoglou2023iris} requirements, as well as Service Level Objectives (SLOs) or Service Level Agreements (SLAs)~\cite{ding2019characterizing}.  
To meet these QoS demands and achieve SLO targets, such models are typically deployed on high-performance Cloud computing platforms, which help mitigate performance bottlenecks while offering scalable, flexible, and cost-effective solutions.  
Modern Cloud systems enhance ML/DL workloads by utilizing distributed computing, specialized hardware accelerators (e.g., GPUs, TPUs)~\cite{wang2019benchmarking}, and optimized networking.  
Many vendors provide Machine Learning as a Service (MLaaS), offering comprehensive solutions for model training, deployment, and inference. Notable platforms include Google Cloud Vertex AI~\cite{mlaas-google-vertex-ai}, AWS SageMaker~\cite{mlaas-aws-sagemaker}, Microsoft Azure ML~\cite{mlaas-azure}, IBM Watson ML~\cite{high2012era}, and Nvidia AI Enterprise~\cite{mlaas-nvidia}.  
However, as service demands continue to surge, Cloud congestion becomes an increasing challenge, leading to limitations in scalability and reliability.  
Additionally, Cloud servers and data centers are typically centralized and located far from end-user devices. 
As a result, latency-sensitive applications often suffer from long round-trip delays, network congestion, and degraded service quality~\cite{grohmann2019monitorless}.  
Furthermore, in data-sensitive sectors such as healthcare, storing sensitive information on third-party infrastructure raises critical privacy and confidentiality concerns~\cite{yang2020data}.

To overcome the limitations of Cloud infrastructures, some of the computational workload is shifted to the Edge, bringing processing closer to where the data is generated.
Specialized AI hardware, such as Nvidia Jetson, Google Edge TPU, and Intel Movidius~\cite{singh2023edge}, along with optimized ML frameworks like ONNX Runtime, TensorFlow, and TensorFlow Lite~\cite{onnx,tensorflow-serving,tflite}, enable efficient AI inference directly on Edge devices. Additionally, a significant amount of modern Edge devices lack dedicated GPUs because of limitations in cost, power consumption, heat dissipation, or physical size. As a result, inference must often be carried out on the CPU, e.g. in mid-range smartphones, wearables, smart home devices and automations.
This advancement allows for low-power, high-performance AI applications, reducing the need for constant Cloud connectivity and shaping the modern paradigm of Edge AI~\cite{letaief2021edge}.  
Additionally, Edge processing improves power efficiency by reducing reliance on energy-intensive Cloud data centers, lowering data transmission costs, and utilizing specialized AI accelerators~\cite{reuther2022ai}.  
However, Edge computing operates with more limited computational and storage resources compared to Cloud-based processing.  
As a result, achieving effective Edge-to-Cloud synergy is crucial for enabling scalable and efficient AI inference across the computing continuum.

A major challenge in achieving seamless Edge-to-Cloud synergy is integrating the QoS guarantees of the Cloud with the efficiency, low overhead, and advantages of the Edge while mitigating the inherent limitations of each processing layer.  
This requires addressing several key challenges:  
\textbf{i) Edge-to-Cloud Scheduling \& Heterogeneity:} Effective collaboration between Edge and Cloud depends on intelligent scheduling of inference workloads while accounting for the diverse architectures of modern processors, such as ARM, RISC-V, and x86. \textbf{ii) Architecture-Aware Inference Deployment:} Many Edge platforms offer configurable power and performance modes (e.g., power modes, turbo boost) that introduce trade-offs between energy efficiency and inference speed. Optimal deployment requires dynamic tuning of execution settings to align with workload demands. Moreover, many devices that operate on the Edge do not have dedicated GPUs due to cost, power, thermal, or size constraints; as a result, inference needs to be performed on the CPU (e.g. mid-range smartphones, wearables, home devices). Furthermore, the growing diversity in CPU architectures—including differences in core counts, instruction sets (e.g., ARM, x86), and power-performance trade-offs further complicates optimized inference scheduling and resource allocation. \textbf{iii) Diverse QoS and SLA Requirements:} Different application domains impose varying QoS constraints. For example, healthcare and automotive applications require millisecond-level inference latency, whereas finance and agriculture can tolerate higher variability, enabling flexible resource allocation across the Edge-to-Cloud continuum.  
\textbf{iv) Inference Engine Variability:} The broad range of inference-serving frameworks, such as TensorFlow~\cite{tensorflow-serving}, ONNX Runtime~\cite{onnx}, and PyTorch~\cite{imambi2021pytorch}, introduces complexity in selecting the optimal model and backend for deployment across different hardware nodes.  
The aforementioned challenges form a multidimensional problem for the efficient deployment of inference-serving workloads within a heterogeneous Edge-to-Cloud continuum.

Efficient inference serving remains a challenging task, as it requires a deep understanding of application architecture, workload characteristics, and the processing capabilities of both Edge and Cloud infrastructures~\cite{ferikoglou2023iris}.  
To address these challenges, prior research has explored inference serving and orchestration solutions specifically tailored for the Edge~\cite{kakolyris2023road,guo2024resource,li2019edge,shi2021dnn}, the Cloud~\cite{ferikoglou2023iris,delimitrou2013paragon,delimitrou2014quasar}, and the broader Edge-to-Cloud continuum~\cite{he2024large,wang2022preemptive,xue2021eosdnn,xue2021ddpqn}.  
SotA research offers effective inference-serving strategies, primarily focusing on QoS-driven performance optimizations~\cite{shahhosseini2022online}, which serve as the core optimization objective of this work.  
However, existing approaches lack architecture-aware scheduling frameworks that dynamically optimize inference serving by adapting to the diverse operating capabilities of heterogeneous Edge and Cloud nodes.  
Additionally, while various scheduling techniques have been proposed for individual layers (Edge or Cloud) and across the continuum, they often fail to fully exploit dynamic resource adaptation and architecture-driven execution modes.  
As a result, these limitations restrict the optimization and customization potential of existing solutions, preventing them from maximizing efficiency under varying workload demands.

In this work, we introduce \textbf{\texttt{SynergAI}}, a novel framework designed for architecture-tuned, performance-efficient inference serving across heterogeneous Edge-to-Cloud nodes. 
The primary optimization goal of \texttt{SynergAI} is to minimize QoS violations for inference engines deployed across the continuum by ensuring efficient, architecture-driven engine placement.
Our solution is based on comprehensive performance and architecture-driven characterization and analysis of discrete inference engines across different Edge/Cloud nodes and operating modes. 
Specifically, we focus on the performance and resource trade-off capabilities of Edge nodes, enabling the development of optimized deployment strategies tailored to each node's unique attributes. The outcome of this analysis provides the crucial tuning parameters for our proposed Edge-to-Cloud orchestrator.
\texttt{SynergAI} utilizes a QoS-aware, architecture-driven scheduling framework that dynamically adjusts the placement of inference jobs based on real-time assessment of QoS violation risks. 
We further evaluate the overhead introduced by our approach and demonstrate that it remains minimal. In addition, we analyze energy consumption, highlighting consistent energy savings across platforms and we present a representative use case that provides an in-depth visualization of \texttt{SynergAI}’s behavior and decision-making process in a real deployment scenario.
To summarize, the key contributions of this work are as follows:

\begin{itemize}  
    \item We conduct an extensive \textbf{characterization and analysis} for performance and architecture-driven tuning and deployment of discrete ML inference engines and models across the Edge-Cloud continuum.  
    
    \item \textbf{We present \texttt{SynergAI}, a novel Edge-to-Cloud scheduling framework} for solving the problem of QoS violations minimization. We incorporate a combination of offline and online mechanisms into our proposed solution, which leverages the characterization and analysis process to perform dynamic task scheduling based on real-time assessments of QoS violation risks. 
    
    \item \textbf{We integrate and evaluate our solution with the Kubernetes framework}, demonstrating that \texttt{SynergAI}'s architecture-driven inference serving enables optimized and architecture-aware deployments on emerging hardware platforms, achieving an average reduction of $2.4\times$ in QoS violations compared to a SotA solution.  
\end{itemize}

The rest of this paper is organized as follows. Section~\ref{sec:related_work} presents an overview of the related work. In Section~\ref{sec:testbed_characterization} we provide an extensive characterization and analysis of discrete ML inference engines and nodes, while in Section~\ref{sec:scheduler} we present \texttt{SynergAI}'s architecture. In Section~\ref{sec:evaluation} the experimental evaluation is presented, as well as an in-depth analysis and discussion. Finally, Section~\ref{sec:conclusion} concludes this research.

%\begin{itemize}
 %   \item Rise of applications utilizing AI/ML. Necessity for effective inference engines
  %  \item Rise of compute, memory and storage requirements and demands.
   % \item Cloud tackles these limitations, through providing "unlimited" resources and high quality services, QoS, etc.
   % \item However Cloud has also some limitations, e.g increased network latency, centralized architecture, etc.
    %\item Close to Edge computing to bring data close to end-user and advantages of the Edge. Show that we still need Cloud.
    %\item Thus, we need an Edge-to-Cloud solution effectively execute inference applications. Consider carefully both the inference engine that we utilize, the target device and the configurations of the device, since they both affect performance and energy. 
    %\item Problem formulation, Edge inference tasks to satisfy QoS, violations. Why not only Edge, not only Cloud and why we need both.    
    %\item State limited works that consider power.
    %\item What we propose? i) Power/Performance-aware (????) scheduling of ML inference serving on the Edge-Cloud continuum, ii) extensive characterization for modern ML inference engines serving for power-aware deployment on the heterogeneous architectures on Edge and Cloud(?), iii) We integrate with Kubernetes, bla bla
    %\item Paper organization
%\end{itemize}

%% file: sections/01_Related_Work.tex
\section{Related Work}
\label{sec:related_work}

In recent years, numerous studies have focused on inference serving and scheduling. 
Systems like TensorFlow Serving~\cite{tensorflow-serving}, TensorFlow Lite~\cite{tflite}, and ONNX Runtime~\cite{onnx} are notable examples of flexible and high-performance serving platforms for ML models, tailored for production environments on both Edge and Cloud infrastructures. 
From an industrial perspective, Nvidia's AI platform offers the Triton Inference Server~\cite{triton-serving}, which facilitates GPU-based inference while also supporting CPU models, though it requires static configuration for model instances. 
Regardless of the setup, effective inference serving relies heavily on efficient scheduling and orchestration. In this section, we present related work, categorized based on the target optimization layer within the Edge-to-Cloud continuum. 
Specifically, we group the examined research into three main categories: (i) \textit{Inference Serving \& Scheduling on the Edge}, (ii) \textit{Inference Serving \& Scheduling on the Cloud}, and (iii) \textit{Inference Serving \& Scheduling across the Edge-Cloud Continuum}.

\textbf{Inference serving \& Scheduling on the Edge:} Several studies have focused on optimizing inference serving at the Edge. 
In~\cite{kakolyris2023road}, the authors propose a DNN partitioning and offloading approach tailored for Edge computing systems. 
Similarly, in~\cite{li2019edge}, a framework is presented that utilizes edge computing for collaborative DNN inference through device-edge synergy. 
In~\cite{shi2021dnn}, a mathematical model is introduced for adaptive DNN model partitioning and inference offloading at the Edge. 
Additionally,~\cite{xiao2022reinforcement} proposes a multi-agent reinforcement learning-based, energy-efficient collaborative inference scheme for Mobile Edge Computing (MEC). 
Lastly, in~\cite{guo2024resource}, a resource-efficient DNN inference method with latency-awareness is proposed, while~\cite{archet2023energy} explores the trade-offs between energy consumption and latency for CNN inference on Edge accelerators.

\textbf{Inference Serving \& Scheduling on the Cloud:} Several studies have investigated inference serving, scheduling, and orchestration at the Cloud layer. 
In~\cite{ferikoglou2023iris}, the authors introduce an interference- and resource-aware predictive orchestrator for ML inference serving. 
In~\cite{delimitrou2013paragon, delimitrou2014quasar}, a scheduling approach for CPU servers is presented, which utilizes load prediction to reduce interference. 
Furthermore,~\cite{tzenetopoulos2022interference} proposes a modular framework that balances incoming workloads based on low-level metrics monitoring. 
Resource partitioning strategies designed to optimize QoS requirements have also been explored in~\cite{patel2020clite}. 
Finally, in~\cite{garefalakis2018m}, workload-specific scheduling techniques are discussed, where different workload classes are handled by distinct schedulers.

\textbf{Inference Serving \& Scheduling across the Edge-Cloud Continuum:} To leverage the advantages of both Edge and Cloud computing, numerous studies have focused on optimizing inference serving and scheduling across the Edge-Cloud continuum. 
In~\cite{he2024large}, the authors propose an active inference-based approach for offloading LLM inference tasks and managing resource allocation in cloud-edge environments. 
The study in~\cite{li2021appealnet} introduces an Edge-Cloud collaborative architecture for DNN inference, while~\cite{wang2022preemptive} presents a preemptive scheduling solution for distributed ML jobs in Edge-Cloud networks. 
Additionally,~\cite{rao2024eco} employs LLMs to dynamically adjust the placement of application tasks across Edge and Cloud layers in response to workload fluctuations. 
In~\cite{xue2021eosdnn}, the authors propose an offloading scheme to accelerate DNN inference in a local-edge-cloud collaborative setting, and~\cite{xue2021ddpqn} presents a deep reinforcement learning-based strategy for optimizing DNN offloading across Edge and Cloud environments.

While several approaches to inference serving and scheduling across the Edge, Cloud, and Edge-Cloud continuum have been explored in existing research, to the best of our knowledge, no study has proposed an architecture-driven scheduling framework that dynamically adapts to the heterogeneous capabilities of Edge and Cloud nodes to optimize inference serving.
Our system takes advantage of the architecture-driven performance features and operating modes of modern Edge/Cloud infrastructures, leveraging their configurable modes to enhance efficiency and minimize QoS violations. 
Through extensive architecture-driven characterization and analysis, our end-to-end orchestrator intelligently allocates ML workloads across the continuum, offering a lightweight and performance-optimized solution.

%% file: sections/02_Characterization.tex
\section{Profiling, Characterization \& Analysis}
\label{sec:testbed_characterization}

\subsection{Inference Serving Testbed}
\label{ssec:testbed}

\textbf{Hardware \& Software Infrastructure}: We conduct our experiments on a high-end dual-socket Intel\textsuperscript{®} Xeon\textsuperscript{®} Gold 5218R (@2.1 GHz) server, equipped with 128 GB of DRAM memory. % I imply that both VMs are deployed on Davinci
On top of this setup, we configure two virtual machines to function as the master (4 vCPUs, 8 GB RAM) and worker (16 vCPUs, 16 GB RAM) nodes of our cluster, respectively, utilizing the KVM hypervisor.  
Additionally, we integrate two more worker nodes on the Edge, which include an Nvidia Jetson AGX (8 CPUs, 32 GB RAM) and an Nvidia Jetson Xavier NX (6 CPUs, 8 GB RAM).
For cluster deployment and orchestration, we leverage Kubernetes~\cite{kubernetes} (v1.28.10) in combination with Containerd (v1.7.2) as the container runtime.

\textbf{Inference Engine Workloads}: For the purposes of this paper, we use the object detection and image classification tasks from the MLPerf Inference benchmark suite~\cite{reddi2020mlperf}.
Table~\ref{tab:mlperf} provides details on the MLPerf inference engines used, including the representation, the model variant, the validation dataset, and the model accuracy.
The accuracy values are solely determined by the model and dataset, and therefore remain consistent across all platforms.
Each inference engine container instance comprises two components: (i) the Inference Engine and (ii) the Load Generator.  
The Inference Engine processes a pre-trained DNN model (e.g., ResNet) using a specified backend framework (e.g., TensorFlow) to perform the designated task.  
Meanwhile, the Load Generator simulates traffic for the Inference Engine and monitors its performance. It takes as input the validation dataset (e.g., ImageNet), the scenario, and the total number of queries to execute.
In our experiments, we employed the Single Stream scenario, where the Load Generator sends one sample per query and waits for execution to complete before dispatching the next one.
Throughout the characterization, we maintain a consistent number of queries in MLPerf across all inference engines to evaluate their performance under the same workload and uncover their internal characteristics. 
This value is the default used in the Single Stream scenario. To eliminate variability caused by network latency, bandwidth limitations, or cloud congestion, all validation datasets used in the experiments are pre-loaded into each target device before execution, respectively.

\textbf{Analysis of Operating Modes in AGX and NX Boards}: The Nvidia AGX and NX boards provide various operating modes to optimize performance, energy efficiency, and thermal management. 
These modes enable the system to switch between peak performance for demanding workloads and lower power consumption for enhanced efficiency in energy-constrained environments. 
Table~\ref{tab:power-modes} presents a detailed overview of the modes analyzed in our research, as derived from the manufacturer's specifications. 
It is important to note that both the AGX and NX boards offer additional modes, but enabling them requires a reboot. 
Since this work focuses on run-time decision making, we avoid using these modes as they introduce overheads which can affect the scheduling system we are developing.
Focusing on the Nvidia Jetson AGX, we observe six available operating modes. 
The CPU frequency ranges from 1200 MHz to 2266 MHz, with the number of active CPUs varying from 2 to 8. 
Additionally, the power budget ranges from 15 Watts to MAXN, generally indicating no fixed upper limit on power consumption. 
This enables the system to operate at peak performance while dynamically adjusting power usage according to workload demands and thermal constraints. 
On the other side, the Nvidia Jetson Xavier NX offers nine available power modes. 
The CPU frequency here ranges from 1200 MHz to 1900 MHz, with 2 to 6 CPUs active, and the power budget ranges from 10 Watts to 20 Watts.

\subsection{Characterizing Inference Serving}
\label{ssec:characterization}

To gain deeper insights into the execution characteristics of the various inference engines listed in Table~\ref{tab:mlperf}, we evaluate the impact of different factors on their performance.
Specifically, we profile the engines to assess how i) x86 and ARM-based workers, ii) vertical resource scaling, and iii) operating modes on ARM-based workers influence their performance. 
At the end of this analysis, we highlight the key insights from our study to determine the optimal modes for implementation in our scheduling system.

\input{tables/mlperf-inference-engines}

\subsubsection{Performance-aware Characterization}

\begin{figure}[h!]
    \centering
    \includegraphics[width=0.25\columnwidth]{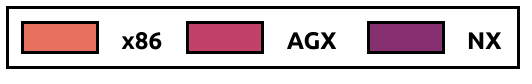}
    \includegraphics[width=\textwidth]{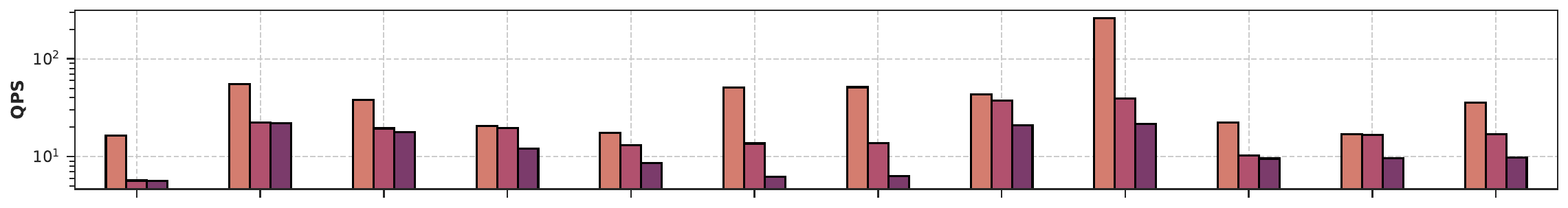}
    \includegraphics[width=\textwidth]{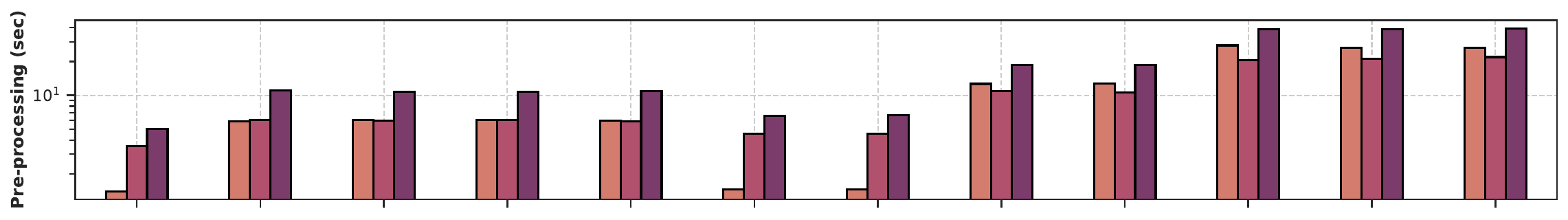}
    \includegraphics[width=\textwidth]{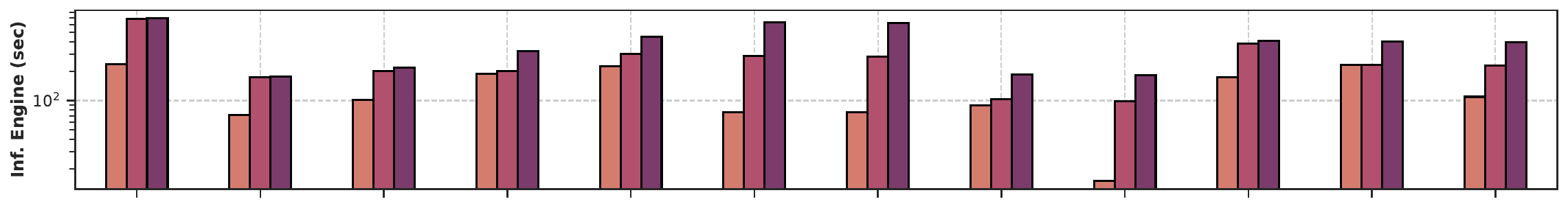}
    \includegraphics[width=\textwidth]{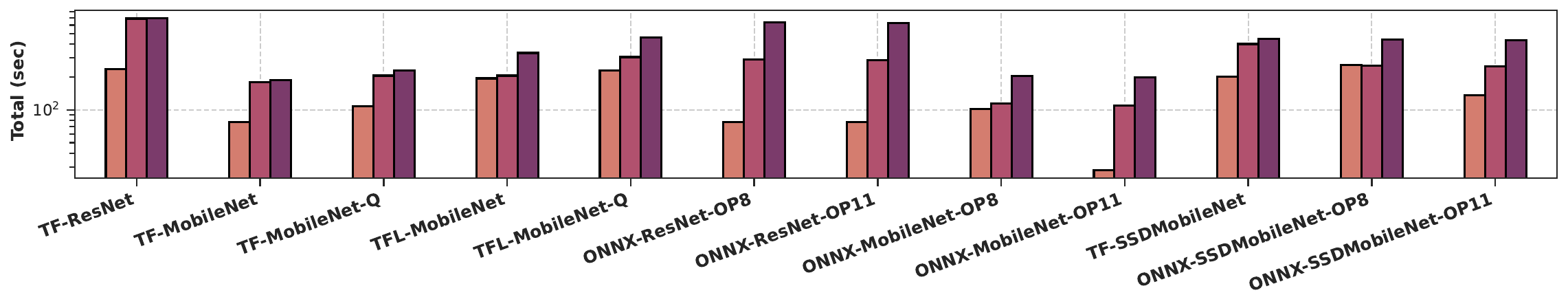}
    \vspace{-20pt}
    \caption{Characterization of the Evaluated Inference Engines Across All Available Workers: QPS (Top), pre-processing time (Upper-middle), inference engine time (Lower-middle) and total execution time (Bottom) in log. scale.}
    \label{fig:Characterization-Default}
\end{figure}

In this section, we profile the inference engines listed in Table~\ref{tab:mlperf} to analyze their performance across all workers in our cluster.
Figure~\ref{fig:Characterization-Default} presents the characterization of the evaluated inference engines across all available workers. 
It illustrates (a) the QPS (Top), (b) the pre-processing time, which includes backend initialization, DNN model loading, and validation data preparation (Upper-middle), (c) the inference engine execution time, which refers to the actual computation (Lower-middle), and (d) the total execution time, as recorded by the Load Generator for each MLPerf Inference Engine, which represents the overall pre-processing and inference engine execution time (Bottom). All metrics are depicted in logarithmic scale. For the x86 worker, QPS values range from 16.5 for TensorFlow ResNet to 259 for ONNX Runtime MobileNet with \texttt{opset-11}, resulting in an average QPS of 52.3.
For total execution time, we observe a distribution ranging from a minimum of 27.8 seconds for ONNX Runtime MobileNet with \texttt{opset-11} to a maximum of 4.3 minutes for ONNX Runtime SSDMobileNet with \texttt{opset-8}.
The average execution time is 2.4 minutes.
An interesting observation is that, despite all inference engines processing the same number of queries, the ONNX ResNet engine with the lowest QPS is not the one with the longest execution time. 
Instead, the shortest execution time is recorded for ONNX Runtime MobileNet with \texttt{opset-11}, which has a QPS of 259.
This behavior is attributed to the preprocessing time, which includes the overhead of initializing the selected backend, setting up the pre-trained DNN model on this backend, and loading the MLPerf validation dataset.
Indeed, TensorFlow ResNet completes this process in just 1.4 seconds, whereas ONNX Runtime MobileNet with \texttt{opset-11} requires $19\times$ longer time for preprocessing.
This significantly impacts performance, ultimately counteracting the advantage of its higher QPS.

For the AGX worker, QPS ranges from 5.7 for TensorFlow ResNet to 39.3 for ONNX Runtime MobileNet with \texttt{opset-11}, with an average of QPS of 19.
The total execution time spans from a minimum of 1.8 minutes for ONNX Runtime MobileNet with \texttt{opset-11} to a maximum of 11.5 minutes for TensorFlow ResNet, averaging 4.6 minutes.  
Similarly, for the NX worker, QPS varies between 5.6 for TensorFlow ResNet and 22 for TensorFlow MobileNet, with an average of 12.5.  
The total execution time ranges from a minimum of 3.1 minutes for TensorFlow MobileNet to a maximum of 11.6 minutes for TensorFlow ResNet, with an average of 6.8 minutes.
For both AGX and NX workers, the inference engine with the highest QPS results in the shortest execution time, while the one with the lowest QPS has the longest execution time, unlike the x86 worker.  
This occurs because preprocessing time increases only slightly, by 10\% on AGX and $1.76\times$ on NX compared to x86.
Preprocessing tasks like image resizing and normalization rely more on memory bandwidth and I/O rather than CPU power, causing minimal slowdown on weaker hardware.  
In contrast, inference is far more computationally intensive, making the increase in preprocessing time negligible compared to the much larger rise in execution time.

\mybox{\noindent $\star$ \textbf{Q1: \textit{How does the performance of inference engines vary when executed on different workers?}}}

The x86 worker outperforms AGX and NX, achieving $2.8\times$ and $4.2\times$ higher QPS, respectively, while also being $2\times$ and $2.8\times$ faster in execution time, respectively.
This is expected, as x86 offers the most powerful CPU and the largest available RAM, followed by AGX and then NX. 
TensorFlow ResNet consistently exhibits the lowest performance across all devices due to its high computational complexity~\cite{limonova2021resnet}, which becomes even more evident on resource-constrained platforms like AGX and NX. 
ONNX MobileNet \texttt{opset-11} is the fastest inference engine on both x86 and AGX, benefiting from ONNX Runtime’s optimizations for efficient parallel execution and low-latency inference. 
However, on NX, TensorFlow MobileNet achieves the best performance.
This analysis highlights the significance of architecture-aware inference deployment, as performance can vary substantially across hardware platforms, even within the same device family, due to the interplay between system architecture, framework optimizations, and the characteristics of the inference model.

$\star$ \textit{\textbf{Key Outcome 1:} Optimal inference engine and model selection varies significantly across different hardware architectures. Inference efficiency is driven by the engine, models and intra-architecture characteristics.} 

\subsubsection{Architecture-Driven Tuning Characterization}

In this section, we examine the impact of various optimizations on the performance of the analyzed inference engines. 
Our approach is divided into two parts: a) for the x86 worker, we evaluate how the number of threads influences performance, and b) for the AGX and NX workers, we investigate how different operating modes affect performance.

\mybox{\noindent $\star$ \textbf{Q2: \textit{How does vertical scaling (i.e., \#Threads) impact performance on x86-based workers?}}}

\begin{figure}[t]
    \centering
    \includegraphics[width=0.35\columnwidth]{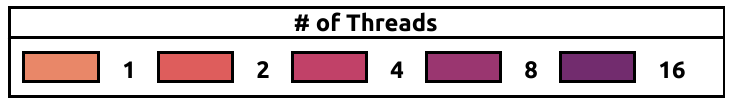}
    \includegraphics[width=\textwidth]{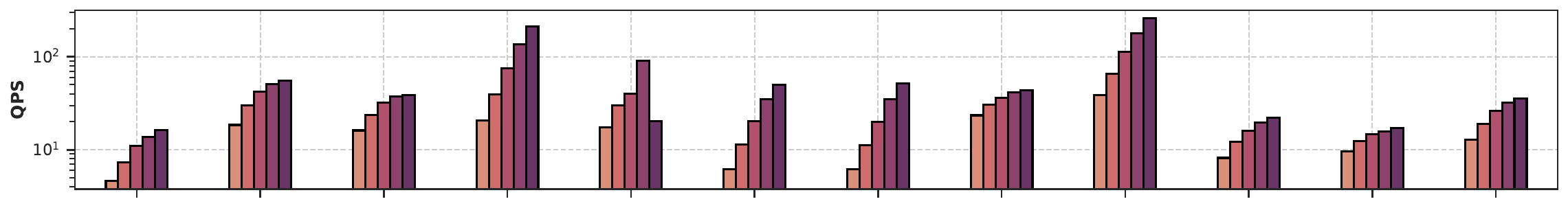}
    \includegraphics[width=\textwidth]{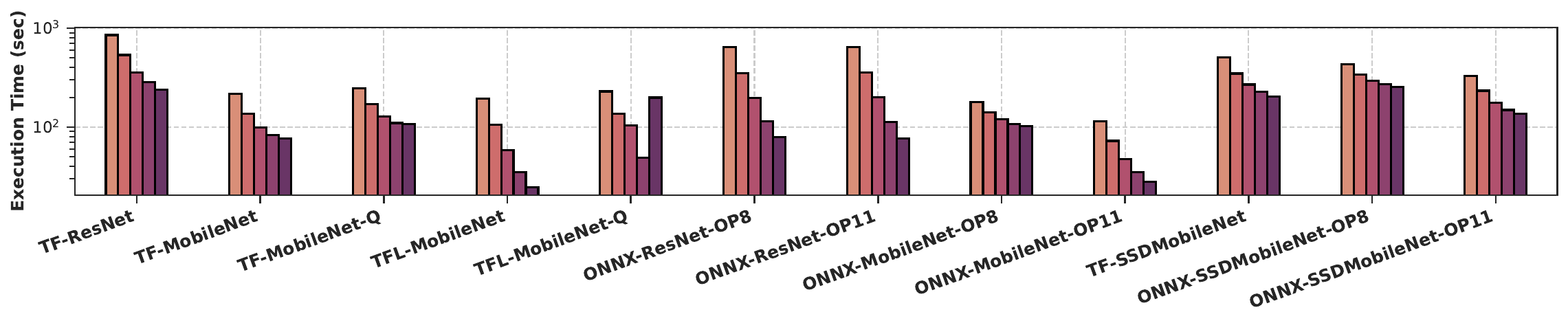}
    \vspace{-20pt}
    \caption{Impact of Thread Scaling on MLPerf Inference Engines in x86 Worker}
    \label{fig:Characterization-Threads}
\end{figure}

In this part of our analysis, we examine the behavior of various inference engines when executed on an x86 worker with different thread counts.
We focus on thread scaling, as it represents one of the most effective approaches for enhancing parallelism and throughput.
Specifically, for inference engines using ONNX Runtime as the backend, we modify the \texttt{INTRA\_OP\_NUM\_THREADS} parameter, and for TensorFlow engines, we adjust the \texttt{INTRA\_OP\_PARALLELISM\_THREADS} setting, which are the most influential parameters, as outlined in prior research~\cite{ferikoglou2023iris}.
Finally, for TensorFlow Lite, we explore the \texttt{NUM\_THREADS} parameter.
Figure~\ref{fig:Characterization-Threads} illustrates the QPS (Top) and execution time (Bottom) across all the inference engines examined (X-axis) for ranging number of threads. 
As shown in Figure~\ref{fig:Characterization-Threads}, increasing the number of available threads results in higher QPS and reduced total execution time. 
Using 2, 4, 8, and 16 threads yields average QPS improvements of $1.6\times$, $2.5\times$, $3.8\times$, and $4.5\times$ respectively, compared to single-threaded execution. 
In terms of total execution time 2, 4, 8, and 16 threads lead to average execution speedups of $1.6\times$, $2.3\times$, $2.9\times$, and $3\times$ compared to single-threaded execution, respectively. 
To quantify the relationship between thread count and performance, we compute the Pearson correlation coefficient~\cite{cohen2009pearson} between execution time and the number of threads. 
Values closer to 1 indicate a stronger linear correlation. 
On average, the correlation is 0.83, suggesting a strong, although not perfectly linear, relationship between thread count and performance.
This observation implies that performance improvements from increasing the number of threads tend to diminish beyond a certain point. 
For example, scaling from 1 to 8 threads yields a speedup of $2.9\times$, while further increasing to 16 threads results in only a marginal improvement to $3\times$. 
These diminishing returns are attributed to factors such as increased synchronization overhead, contention for shared resources (e.g., cache and memory bandwidth), and inefficiencies in parallel execution. Finally, preprocessing time shows limited responsiveness to increases in thread count, resulting in a total execution time reduction of only 21\% at 16 threads compared to single-threaded execution. This modest improvement is expected, as we previously mentioned that preprocessing tasks are more dependent on memory bandwidth and I/O operations than on CPU parallelism.
As a result, fully utilizing all available threads does not guarantee proportional performance gains, indicating that near-optimal performance is often achievable with a reduced number of threads.

This effect becomes more noticeable in inference engines with relatively short total execution times. 
As the number of threads increases and the inference engine becomes faster, the preprocessing stage emerges as the primary bottleneck. 
For instance, in the ONNX Runtime implementation of SSD MobileNet with \texttt{opset-8}, increasing the thread count beyond 4 causes preprocessing to dominate the overall execution time, thereby reducing the benefit of additional threads.
In addition, the scalability of performance is significantly influenced by the choice of backend and quantization strategy. 
In the case of TensorFlow Lite's quantized MobileNet model, performance improves as expected when scaling from 1 to 8 threads. 
However, at 16 threads, both queries per second (QPS) and execution time remain nearly unchanged compared to single-threaded execution. 
This behavior is primarily driven by two factors. 
First, TensorFlow Lite's threading model is optimized for embedded devices and is designed to prioritize low-latency execution rather than high degrees of parallelism. 
As a result, it does not effectively distribute workloads across a large number of threads on x86 CPUs, which are not its primary target architecture. 
Consequently, performance improvements tend to plateau beyond 8 threads.
Second, quantization itself plays a crucial role in determining threading efficiency. 
\texttt{int8} operations are typically faster than floating-point operations due to their lower computational complexity and reduced memory bandwidth requirements~\cite{van2023fp8,int8fp32}. 
However, because these operations are inherently lightweight and complete rapidly, the overhead associated with coordinating a larger number of threads can negate the expected performance gains. 
In some cases, excessive threading may even degrade performance due to increased synchronization overhead.

$\star$ \textit{\textbf{Key Outcome 2:} Increasing the number of threads on x86-based workers enhances inference performance, but the improvements taper off beyond a certain point. This suggests that near-optimal performance can be achieved without fully utilizing all available threads, allowing for more efficient resource usage.}

% $\star$ \textit{\textbf{Key Outcome 2:} Increasing the number of threads improves inference performance on x86-based workers, but with diminishing returns beyond a specific number of threads, suggesting that near-optimal performance can be achieved without utilizing all available threads.}

% $\star$ \textit{\textbf{Key Outcome 3:} Performance does not scale fully linearly with the number of threads due to contention, synchronization overheads, and workload-specific limitations, leading to diminishing returns beyond a certain threshold. Thus, performance over-consumption can be avoided.}

\mybox{\noindent $\star$ \textbf{Q3: \textit{How do operating modes affect ARM-based workers?}}}

\input{tables/power-modes-new}

\begin{figure*}[t]
    \captionsetup[subfigure]{labelformat=empty}
    
    % First row of subfigures
    \begin{minipage}[t]{\textwidth}
        \centering
        \subfloat[]{
            \includegraphics[width=.47\columnwidth]{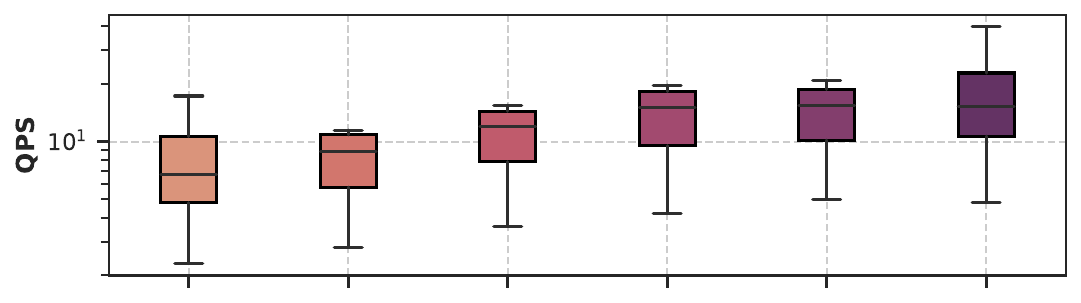}
        } \hspace*{\fill}
        \subfloat[]{
            \includegraphics[width=.47\columnwidth]{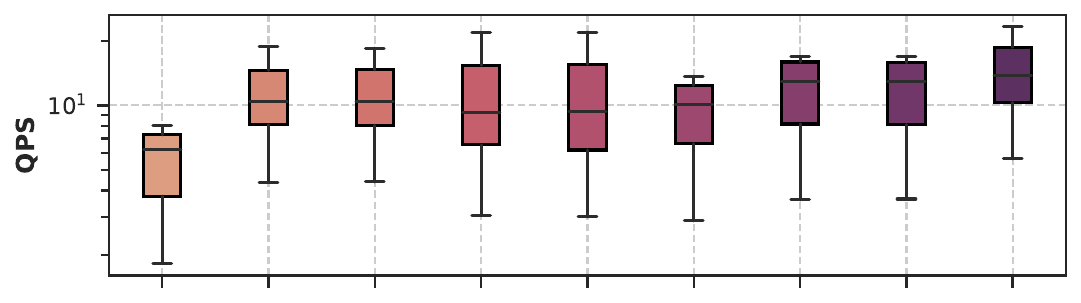}
        }
        \vspace{-25pt}
    \end{minipage}
    % Second row of subfigures
    \begin{minipage}[t]{\textwidth}
        \centering
        \subfloat[(a) AGX]{
            \includegraphics[width=.47\columnwidth]{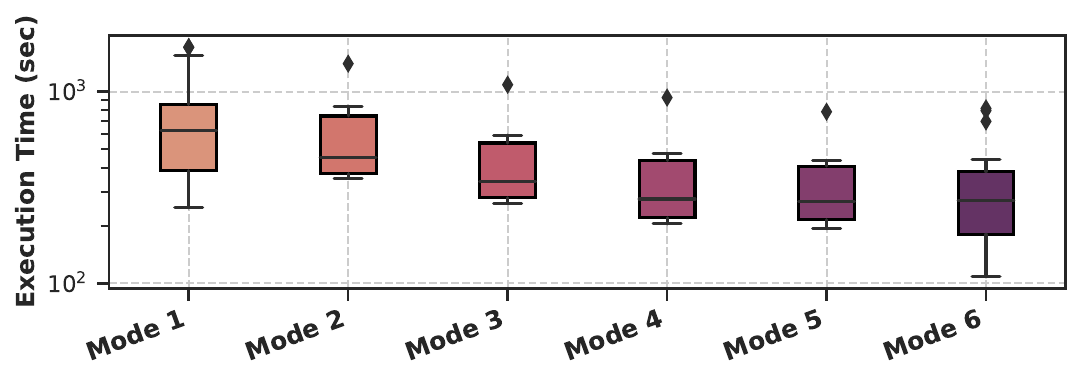}
        } \hspace*{\fill}
        \subfloat[(b) NX]{
            \includegraphics[width=.47\columnwidth]{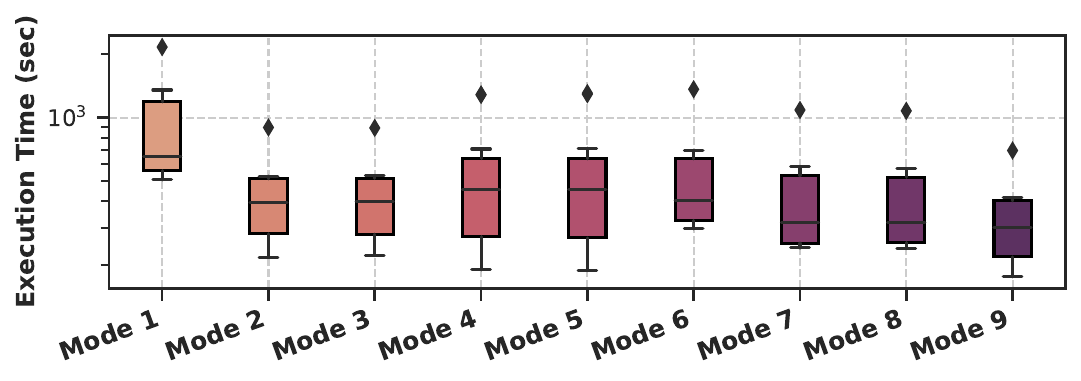}
        }
    \end{minipage}
    \vspace{-10pt}
    \caption{Impact of Operating Modes on the QPS (Top) and execution time (Bottom) for the MLPerf Inference Engines for AGX and NX}
    \label{fig:AGX_NX_Power_Modes}
\end{figure*}

In this part of our analysis, we assess the impact of operating modes on ARM-based workers. 
Rather than directly analyzing thread scaling as done on the x86 worker, we evaluate its effect through operating modes, which simultaneously modify parameters such as CPU frequency, core availability, and power allocation, all of which influence performance.
Figure~\ref{fig:AGX_NX_Power_Modes} illustrates the distributions of QPS and total execution time for the AGX and NX boards across the available operating modes, as outlined in Table~\ref{tab:power-modes}.
Focusing on the AGX worker, we observe that Mode 6 stands out with the highest QPS and, consequently, the lowest execution time. 
The QPS distribution ranges from 4.8 to 39.8, averaging at 18.4, while execution time ranges from 1.8 to 13.6 minutes, with an average of 5.1 minutes. 
Specifically, Mode 6 delivers a QPS that is $1.3\times$ higher than Mode 5, $1.4\times$ higher than Mode 4, $1.8\times$ higher than Mode 3, $2\times$ higher than Mode 2, and $2.2\times$ higher than Mode 1. 
In contrast, the slowest mode, Mode 1, shows a QPS distribution from 4.8 to 17.3, averaging 8.2, with execution times ranging from 4.1 to 28.5 minutes, averaging 11.4 minutes.
Mode 6 is expected to perform the best due to its highest maximum operating frequency, utilization of all available CPUs, and lack of power budget limitations. 
On the other hand, although Mode 1 offers more online CPUs and a higher power budget, its lower frequency results in it being the slowest option.

For the NX worker, Mode 9 achieves the highest QPS.  
The QPS distribution ranges from 5.6 to 23.3, averaging 14.3, while execution time varies between 2.9 and 11.6 minutes, with an average of 5.8 minutes.  
Specifically, Mode 9 delivers a QPS that is $1.2\times$ higher than Modes 7 and 8, $1.3\times$ higher than Modes 2, 3, 4, and 5, $1.5\times$ higher than Mode 6, and $2.5\times$ higher than Mode 1.  
The lowest-performing mode is Mode 1, with a QPS distribution ranging from 1.8 to 8.1, averaging 5.5, and execution times between 8.4 and 36 minutes, with an average of 16.9 minutes.  
Mode 9 achieves high performance by utilizing the highest CPU frequency while using only four threads and operating at the lowest power budget.  
In contrast, Mode 1, which has the lowest maximum CPU frequency of 1200 MHz, along with the same number of CPUs and power budget, results in the slowest execution.

$\star$ \textit{\textbf{Key Outcome 3:} Operating modes significantly impact performance on ARM-based workers, with higher CPU frequencies leading to better QPS and lower execution times.} 

\mybox{\noindent $\star$ \textbf{Q4: \textit{Which parameters (e.g., \#CPUs, frequency) have the greatest impact on performance?}}}

\begin{figure*}[t]
    \captionsetup[subfigure]{labelformat=empty}

    \begin{minipage}[t]{\textwidth}
        \centering
        \subfloat[]{
            \includegraphics[width=.31\columnwidth]{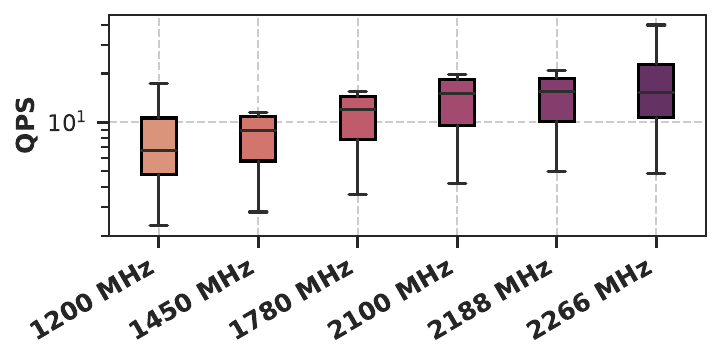}
        }
        \subfloat[]{
            \includegraphics[width=.31\columnwidth]{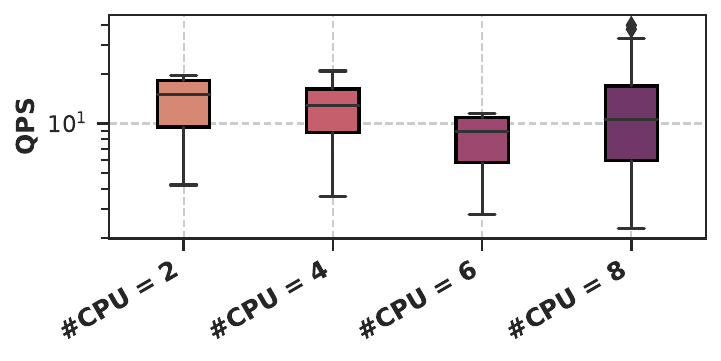}
        }
        \subfloat[]{
            \includegraphics[width=.31\columnwidth]{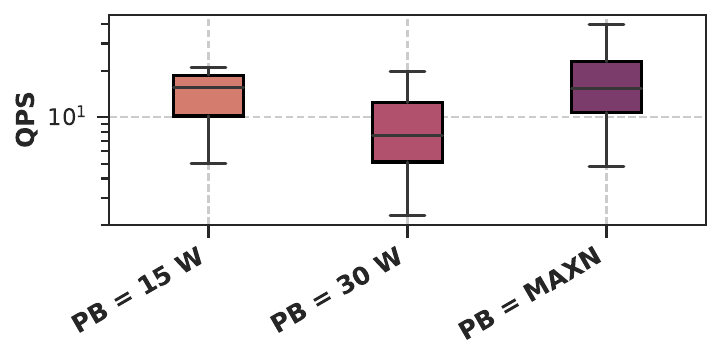}
        }
    \end{minipage}
    
    \vspace{-20pt}
    
    \caption{Impact of Frequency, \#CPUs and Power Budget on Performance for AGX Worker}
    \label{fig:AGX_Impact}
\end{figure*}

\begin{figure*}[t]
    \captionsetup[subfigure]{labelformat=empty}

    \begin{minipage}[t]{\textwidth}
        \centering
        \subfloat[]{
            \includegraphics[width=.31\columnwidth]{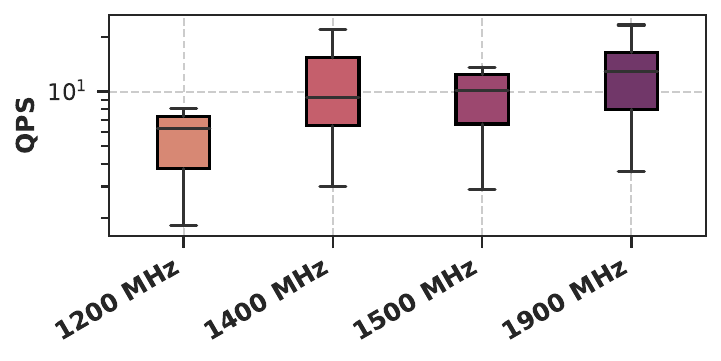}
        }
        \subfloat[]{
            \includegraphics[width=.31\columnwidth]{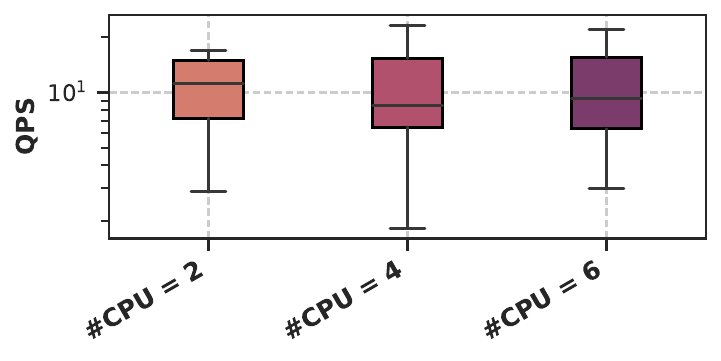}
        }
        \subfloat[]{
            \includegraphics[width=.31\columnwidth]{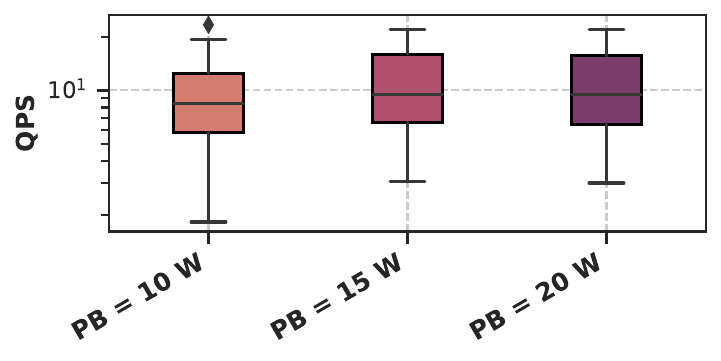}
        }
    \end{minipage}
    
    \vspace{-20pt}
    
    \caption{Impact of Frequency, \#CPUs and Power Budget on Performance for NX Worker}
    \label{fig:NX_Impact}
\end{figure*}

To gain a deeper understanding of why the aforementioned operating modes are significant, we analyze how the performance of the inference engines is affected by the key aspects of each operating mode for both the AGX and NX workers. 
Figures~\ref{fig:AGX_Impact} and~\ref{fig:NX_Impact} show the distributions of QPS across different maximum operating clock frequencies (Left), the number of online CPUs (Center), and the available power budget (Right).
As observed for both AGX and NX workers, increasing the maximum CPU frequency results in higher performance.
For the AGX worker, raising the CPU frequency from 1200 MHz to 1450 MHz leads to an average $2.5\%$ increase in QPS. Increasing to 1780 MHz results in a $1.35\times$ boost, to 2100 MHz a $1.7\times$ increase, to 2188 MHz a $1.8\times$ increase, and to 2266 MHz a $2.3\times$ improvement on average.
A similar behavior is observed for the NX worker, where moving from the same initial CPU frequency to the next available frequency results in a $1.7\times$, $2\times$, and $2.3\times$ higher QPS on average.

Regarding the number of available online CPUs, we observe a counterintuitive trend.  
For the AGX worker, increasing the number of online CPUs generally leads to a decrease in QPS on average. 
Specifically, utilizing 4 online CPUs results in an 8\% decrease in QPS, while having 6 online CPUs leads to a $1.7\times$ lower QPS compared to using 2 online CPUs.
With 8 online CPUs, QPS remains nearly the same as with 2 CPUs, showing only a slight 1\% drop on average.  
For the NX worker, QPS remains relatively stable. 
We observe a slight 2\% decrease on average with 4 CPUs, followed by a 4\% increase when using 6 CPUs compared to the 2-CPU configuration.
To understand this behavior, we focus on the 2-CPU operating modes. 
On the AGX worker, the 2-CPU configuration corresponds to Mode 4, which operates at a high frequency of 2100 MHz. 
In contrast, the 6-CPU configuration, where we observe the largest QPS drop, only includes Mode 2, which has a lower frequency of 1450 MHz. 
Even with 8 online CPUs, which include the best-performing Mode 6, the presence of Mode 1 operating at a lower frequency of 1200 MHz causes a decline in the QPS distribution. 
Similar observations hold for the NX worker.  
This suggests that operating frequency is a more critical factor than the number of online CPUs. 
If performance is the primary objective, it is often preferable to choose an operating mode with fewer CPUs but a higher frequency, rather than increasing the number of CPUs at the cost of reduced frequency.

As for the available power budget of the operating modes for the AGX worker, the MAXN power budget stands out with an average QPS of 18.4. 
The next best power budget is 15 Watts, which results in a $1.3\times$ decrease, followed by 30 Watts, which leads to a $2\times$ decrease compared to MAXN.  
Although one might expect that the 30-Watt mode would perform better than the 15-Watt mode, a deeper analysis of the operating modes reveals that the 15-Watt power budget is exclusively associated with Mode 6, which has the second-highest operating frequency. 
In contrast, the 30-Watt power budget includes multiple modes with significantly lower frequencies, leading to a lower average QPS.  
For the NX worker, the behavior follows expectations, with the lowest QPS values observed at 10 Watts, averaging 9.7 QPS. 
At 15 Watts and 20 Watts, the QPS values are similar, averaging 11.3. 
The similarity in performance between 15 Watts and 20 Watts is attributed to the wider range of operating frequencies available within these power budgets, preventing any single mode from standing out.

$\star$ \textit{\textbf{Key Outcome 4:} CPU frequency has the greatest impact on performance, outweighing the number of online CPUs, while power budget influences performance indirectly based on the frequency and modes it enables.}

To summarize, our evaluation shows that optimal inference performance is closely tied to the hardware characteristics of the target architecture. 
The selection of inference engine and model significantly impacts performance and varies across architectures due to engine-specific optimizations and architectural differences. 
On x86-based workers, increasing the number of threads improves performance, but benefits diminish beyond 8 threads, making it a practical choice for balancing efficiency and resource usage. 
In contrast, ARM-based workers are highly sensitive to operating modes, with CPU frequency being the dominant factor driving throughput and execution time. 
Other parameters, such as core count and power budget, have a more indirect and comparatively minor influence on performance.

% \MK{Add here a paragraph showing that the outcome is XXXX, and write what we do on : i) new device, based on the observations we select the max Freq. mode (stress that it is not identical with the max threads), ii) new model send it on its optimal mapping/mode. Say that in parallel we can perform DSE for the new model/device}
% \MK{Can we show maybe here which is the "optimal" mode(s) per board? and claim that these will be used on the next section, e.g: AGX: The optimal configuration for performance is MAXN (higher power budget, higher CPU frequency) combined with fewer online CPUs and NX: The optimal configuration also involves a balance of higher CPU frequencies with fewer CPUs, and similar performance is achieved across 15W and 20W power settings. This will form a better bridge with the next Section and a smooth transition}

%% file: tables/mlperf-inference-engines.tex
\setlength\extrarowheight{2pt}
\begin{table}[b]
\footnotesize
\caption{Supported MLPerf Inference Engines, Model Variant, Dataset and Corresponding Accuracy}
\centering
\resizebox{\columnwidth}{!}{%
\begin{tabular}{p{0.28\columnwidth}|p{0.25\columnwidth}|p{0.3\columnwidth}|p{0.12\columnwidth}|p{0.12\columnwidth}}
\specialrule{1.5pt}{0pt}{0pt}
\cellcolor{gray!10}\textbf{Task} & \cellcolor{gray!10}{\textbf{Representation}} & \cellcolor{gray!10}\textbf{Model Variant} & \cellcolor{gray!10}\textbf{Dataset} & \cellcolor{gray!10}\textbf{Accuracy}\\
\specialrule{1.5pt}{0pt}{0pt}
 & & ResNet50 & ImageNet & 76.456\% \\
\cline{3-5}
 & \multirow{1}{*}{\textbf{Tensorflow (TF)~\cite{tensorflow-serving}}} & MobileNet & ImageNet & 71.676\% \\
\cline{3-5}
 &  & MobileNet Quantized (Q) & ImageNet & 70.694\% \\
\cline{2-5}
 &  & MobileNet & ImageNet & 71.676\% \\
\cline{3-5}
\multirow{-1}{*}{{\textbf{Image Classification}}} & \multirow{-2}{*}{\textbf{TFLite (TFL)~\cite{tflite}}} & MobileNet Quantized & ImageNet & 70.762\% \\
\cline{2-5}
 & \multirow{4}{*}{\textbf{ONNX Runtime~\cite{onnx}}} & ResNet50 \texttt{opset-8} (OP8) & ImageNet & 76.456\% \\
\cline{3-5}
 & & ResNet50 \texttt{opset-11} (OP11) & ImageNet & 76.456\% \\
\cline{3-5}
 &  & MobileNet \texttt{opset-8} & ImageNet & 71.676\% \\
\cline{3-5}
 &  & MobileNet \texttt{opset-11} & ImageNet & 71.676\% \\
\specialrule{1.5pt}{0pt}{0pt}

 & \textbf{TensorFlow~\cite{tensorflow-serving}} & SSDMobileNet & Coco 300 & mAP 0.234\\
\cline{2-5}
 & & SSDMobileNet \texttt{opset-8} & Coco 300 & mAP 0.23\\
\cline{3-5}
\multirow{-3}{*}{{\textbf{Object Detection}}}
 & \multirow{-2}{*}{\textbf{ONNX Runtime~\cite{onnx}}} & SSDMobileNet \texttt{opset-11} & Coco 300 & mAP 0.23\\
\specialrule{1.5pt}{0pt}{0pt}
\end{tabular}
}
\label{tab:mlperf}
\end{table}

%% file: tables/power-modes-new.tex
\setlength\extrarowheight{2pt}
\begin{table}[b]
\footnotesize
\caption{Power Mode Configurations for Nvidia Jetson AGX and Xavier NX Boards}
\centering
\resizebox{\columnwidth}{!}{%
\begin{tabular}{p{0.22\columnwidth}|p{0.10\columnwidth}|p{0.25\columnwidth}|p{0.15\columnwidth}|p{0.18\columnwidth}}
\specialrule{1.5pt}{0pt}{0pt}
\cellcolor{gray!10}\textbf{Board} & \cellcolor{gray!10}\textbf{Mode} & \cellcolor{gray!10}\textbf{Max CPU Frequency (MHz)} & \cellcolor{gray!10}\textbf{Online CPUs} &
\cellcolor{gray!10}\textbf{Power Budget (W)} \\
\specialrule{1.5pt}{0pt}{0pt}

\multirow{6}{*}{\textbf{Nvidia Jetson AGX}}  
    & Mode 1 & 1200 & 8 & 30 \\
    \cline{2-5}
    & Mode 2 & 1450 & 6 & 30 \\
    \cline{2-5}
    & Mode 3 & 1780 & 4 & 30 \\
    \cline{2-5}
    & Mode 4 & 2100 & 2 & 30 \\
    \cline{2-5}
    & Mode 5 & 2188 & 4 & 15 \\
    \cline{2-5}
    & Mode 6 & 2266 & 8 & MAXN \\
    \cline{2-5}
\specialrule{1.5pt}{0pt}{0pt}

\multirow{8}{*}{\textbf{Nvidia Jetson Xavier NX}}  
    & Mode 1 & 1200 & 4 & 10 \\
    \cline{2-5}
    & Mode 2 & 1400 & 4 & 15 \\
    \cline{2-5}
    & Mode 3 & 1400 & 4 & 20 \\
    \cline{2-5}
    & Mode 4 & 1400 & 6 & 15 \\
    \cline{2-5}
    & Mode 5 & 1400 & 6 & 20 \\
    \cline{2-5}
    & Mode 6 & 1500 & 2 & 10 \\
    \cline{2-5}
    & Mode 7 & 1900 & 2 & 15 \\
    \cline{2-5}
    & Mode 8 & 1900 & 2 & 20 \\
    \cline{2-5}
    & Mode 9 & 1900 & 4 & 10 \\
    \cline{2-5}
\specialrule{1.5pt}{0pt}{0pt}
\end{tabular}
}
\label{tab:power-modes}
\end{table}

%% file: sections/03_Scheduler.tex
\section{SynergAI Scheduling Framework}
\label{sec:scheduler}

\begin{figure*}[t]
    \centering
    \resizebox{0.985\textwidth}{!}{
    \includegraphics[width=\columnwidth]{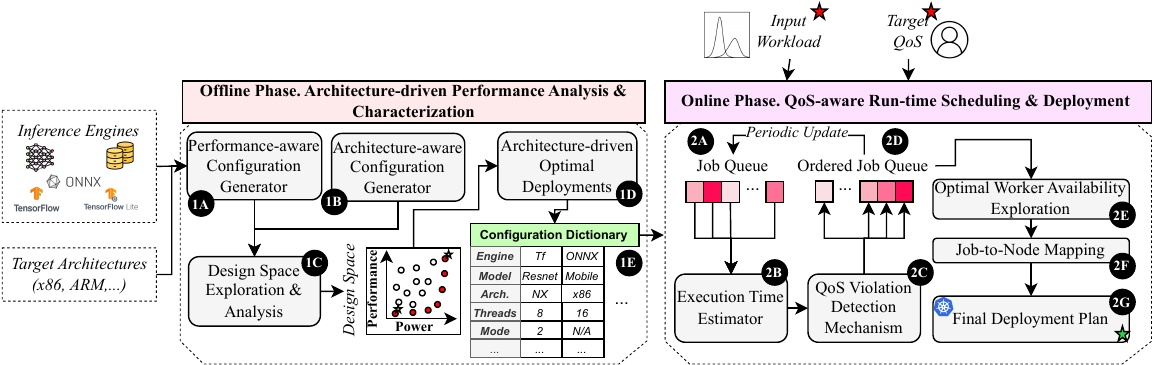}}
    \caption{Overview of the Offline and Online Phases of the \texttt{SynergAI} Framework}
    \label{fig:methodology}
\end{figure*}

Based on the insights derived by the characterization and analysis presented in Section~\ref{ssec:characterization}, we design \texttt{SynergAI}. 
Its main goal is to satisfy the QoS requirements of the deployed inference engines by reducing the number of QoS violations.
\texttt{SynergAI} is illustrated in Figure~\ref{fig:methodology} and aims to provide a synergetic Edge-to-Cloud solution and consists of two discrete phases, i.e. i) \textit{Architecture-driven Performance Analysis \&
Characterization (Offline)} detailed in Sec.~\ref{ssec:offline} and ii) \textit{QoS-aware Run-time Scheduling \& Deployment (Online)}, analyzed in Sec.~\ref{ssec:online}.

\subsection{Offline Phase: Architecture-driven Performance Analysis \&
Characterization}
\label{ssec:offline}

% \MK{Say how we handle an unknown task/device. First we deploy based on the outcomes of the profiling, i.e. max freq. mode with min power. Say that in parallel we can perform DSE for the new model/device to fine-tune if there are special characteristics.}

The offline phase aims to evaluate how each inference engine performs under various architecture-specific optimizations, driven by the characterization and analysis presented in Sec.~\ref{ssec:characterization}. 
As input we provide the target inference engines intended for deployment within the proposed inference serving system, as well as the target Edge/Cloud nodes.
More specifically, the inference engine consists of the backend (e.g., TensorFlow, ONNX Runtime), the pre-trained DNN model (e.g., ResNet), and the validation dataset (e.g., ImageNet), while the target nodes are characterized by their underlying architecture of workers (e.g., x86, ARM) and their operating modes.
The result of the offline phase is the creation of a configuration dictionary for the discrete inference engines, which encapsulates the data to be used later for job scheduling during the Online phase (Sec.~\ref{ssec:online}).

Initially, the \textit{Performance-aware Configuration Generator}~\circled{1A} examines the available levels of parallelism for each architectural configuration, optimizing execution efficiency accordingly. 
This process involves configuring the degree of vertical scaling for each backend within the inference engine on each worker node, as highlighted in prior research in~\cite{ferikoglou2023iris}. 
By systematically exploring parallelism options, the system ensures that inference workloads are optimally distributed across computational resources, maximizing throughput and minimizing latency. 
In addition to parallelism optimization, \texttt{SynergAI} integrates the \textit{Architecture-aware Configuration Generator}~\circled{1B}, designed to enhance efficiency in architectures where operating mode tuning is feasible. 
This module integrates the discrete operating modes available on specific hardware platforms, such as the AGX and NX boards, to assess their impact on performance within discrete power budgets. 
By dynamically selecting the most efficient operating mode configuration, \texttt{SynergAI} ensures that inference workloads are executed with an optimal balance between computational speed, threading, and efficiency, leading to architecture-tuned deployments. 
This dual optimization strategy—leveraging both parallelism exploration and architecture-aware tuning—enables \texttt{SynergAI} to adaptively configure heterogeneous Edge and Cloud nodes, optimizing resource utilization while meeting QoS constraints.

Once all potential configurations for an inference engine across the available workers are collected, a thorough \textit{Design Space Exploration \& Analysis}~\circled{1C} is performed. 
This step involves systematically evaluating each configuration to identify the most efficient and high-performing settings for different workers, considering the specific operating modes. 
By carefully exploring these configurations, we gain valuable insights into the trade-offs between speed and efficiency, ultimately resulting in a well-balanced and optimized deployment strategy.
During this exploration, we gather detailed performance metrics, such as the achieved QPS, total execution time, pre-processing time, actual computation time, threading, and more. 
These metrics provide a comprehensive view of the inference engine's performance under varying configurations. 
Following an in-depth analysis of the collected data, we identify the \textit{Architecture-driven Optimal Deployments}~\circled{1D} for each inference engine, selecting the best-performing configurations from all available architectures. 
This data is then stored in a structured database, ensuring easy access and retrieval when required.
The final dataset forms the \textit{Configuration Dictionary}~\circled{1E}, which serves as a reference for the \texttt{SynergAI} orchestrator. 
This dictionary includes crucial elements for optimal inference serving, such as the model, backend, operating mode, and other system configurations. 
It acts as the foundation for job scheduling in the \textit{Online Phase} (Sec.~\ref{ssec:online}), enabling the orchestrator to make intelligent, architecture-tuned decisions that maximize the efficiency of the entire Edge-Cloud system.

\subsection{Online Phase: QoS-aware Run-time Scheduling \& Deployment}
\label{ssec:online}

\input{tables/maths_description}

During the Online Phase, \texttt{SynergAI} continuously processes incoming inference workloads while ensuring compliance with the target QoS objectives. 
The scheduling policy constructs a \textit{Job Queue}~\circled{2A}, where each job represents an inference engine with specific execution requirements, including the total number of queries to be processed and the target QoS. 
Let \( J \) be the set of incoming jobs, defined as \( J = \{j_1, j_2, \dots, j_N\} \), and \( W \) be the set of workers, defined as \( W = \{w_1, w_2, \dots, w_M\} \).
The formulation of our target problem and decision-making process is presented in Eq.~\ref{eq:remaining_time}-Eq.~\ref{eq:optimal_worker_selection}, where each equation is evaluated for every job \( j \in J \). 
The key parameters are briefly explained in Table~\ref{tab:variables}.

\begin{equation} 
T_{\text{Remaining}, j} = T_{\text{QoS}, j} - T_{\text{Waiting}, j}, \quad \forall j \in J
\label{eq:remaining_time} 
\end{equation}

\begin{equation} 
   T_{\text{Estimated}, j, w} = T_{\text{Pre-processing}, j} + \frac{q}{QPS^{c^*_{j, w}}}, \quad \forall j \in J, \forall w \in W
   \label{eq:execution_time_estimation}
\end{equation}  

\begin{equation}  
W_{\text{acceptable}, j} = \{w \in W \mid T_{\text{Remaining}, j} \geq T_{\text{Estimated}, j, w} \}, \quad \forall j \in J
\label{eq:acceptable_workers}  
\end{equation}

\begin{equation}  
w^*_{j} = \arg \min_{w \in W_{\text{acceptable}, j}} T_{\text{Estimated}, j, w}
\label{eq:optimal_worker_selection}  
\end{equation}

Let \( T_{\text{QoS}, j} \) denote the time specified by the user for executing job \( j \), and let \( T_{\text{Waiting}, j} \) represent the time elapsed since the job \( j \) was submitted to the queue.  
We define the remaining time before a QoS violation for \( j \) as \( T_{\text{Remaining}, j} \), expressed in Eq.~\ref{eq:remaining_time}.  
As \( T_{\text{Waiting}, j} \) increases, \( T_{\text{Remaining}, j} \) decreases, approaching zero, indicating an increasing urgency for execution.  
To effectively manage job prioritization and prevent QoS violations, \texttt{SynergAI} continuously monitors these time constraints. 
Subsequently, the queued jobs are forwarded to the \textit{Execution Time Estimator}~\circled{2B}, which is the key enabler of adaptive scheduling, allowing \texttt{SynergAI} to dynamically determine job execution order.  
Utilizing the \textit{Configuration Dictionary}~\circled{1D}, \texttt{SynergAI} selects the optimal configuration for each worker that maximizes QPS for a given inference engine, denoted as \( c^*_{j, w} \) for each job \( j \) and worker \( w \).  
Given a request to execute \( q \) queries and the profiled pre-processing time for job \( j \), denoted as \( T_{\text{Pre-processing}, j} \), the estimated execution time \( T_{\text{Estimated}, j, w} \) is formulated in Eq.~\ref{eq:execution_time_estimation}, where \( QPS^{c^*_{j, w}} \) represents the QPS achieved by configuration \( c^*_{j, w} \). 
This information is incorporated into the \textit{QoS Violation Detection Mechanism}~\circled{2C}, whose goal is to identify jobs at risk of exceeding their QoS constraints, determine the optimal worker for deployment, and prioritize jobs with a higher probability of violation.  
The system evaluates the urgency of each job by computing the difference between \( T_{\text{Remaining}, j} \) and \( T_{\text{Estimated}, j, w} \) for each worker \( w \).  
As the difference approaches zero, urgency increases, while a negative value indicates an inevitable QoS violation.

The final worker selection is determined through a set of acceptable workers, which can guarantee QoS compliance given the remaining time \( T_{\text{Remaining}, j} \), as defined in Eq.~\ref{eq:acceptable_workers}.  
These workers are then sorted in ascending order by their estimated execution time.  
The optimal worker \( w^*_{j} \) for the given job is the worker that minimizes the estimated execution time \( T_{\text{Estimated}, j, w} \), as formulated in Eq.~\ref{eq:optimal_worker_selection}.  
This ensures that jobs are assigned to the fastest available worker nodes, when not occupied, thereby minimizing execution time while meeting QoS requirements.
By maintaining this prioritized list of suitable nodes for each job, we address scenarios where the optimal worker is currently occupied. 
In such cases, the scheduler can immediately assign the job to the next best available worker, reducing waiting time.
Once urgency values for all jobs are computed and the optimal assignments are identified, the queue is sorted in descending order based on urgency, ensuring that jobs at the highest risk of QoS violations are prioritized, composing the \textit{Ordered Job Queue}~\circled{2D}.  
If no worker can complete a job within its required execution time (i.e., a QoS violation occurs), the job is de-prioritized and moved to the end of the queue, allowing higher-priority jobs with active QoS constraints to be processed first. 
To ensure continuous optimization, a periodic update mechanism reassesses all jobs in the queue, mapping them to the most suitable worker based on updated waiting times and dynamically reordering the queue as needed.
Jobs are dequeued sequentially, each associated with a list of (\( w, c^*_{j, w} \)) pairs, where each pair consists of a worker and its optimal configuration for the given inference engine.  
Finally, when a job is dequeued for execution, \textit{Optimal Worker Availability Exploration}~\circled{2E} iterates through the sorted set \( W_{\text{acceptable}, j} \), checking each worker's availability until it finds the first available worker:  
(\( w_{j} \), \( c^*_{j, w^*_{j}} \)), starting from the optimal worker \( w^*_{j} \). 
If no workers are available, the job waits until one becomes free.
Once a suitable worker and configuration are identified, the scheduler proceeds with the final \textit{Job-to-Node Mapping}~\circled{2F}.
Once the job is mapped to the corresponding node, the \textit{Final Deployment Plan}~\circled{2G} is derived accordingly and deployed via Kubernetes.  
This iterative process continues until all jobs are scheduled and executed successfully within the Edge/Cloud cluster.  

\textit{\textbf{Incorporating new devices and inference engines}}: The proposed Online methodology operates based on pre-characterized inference engines and worker nodes. 
The methodology is also capable of accommodating newly added devices or inference engines. 
In such cases, we leverage the insights from Section~\ref{ssec:characterization} to guide their deployment, even in the absence of a complete configuration profile.
For new workers, if multiple frequency settings are available, we choose the configuration with the highest frequency, as it generally provides the best performance. 
Among configurations with similar frequencies, we prioritize those with the second-highest number of CPU cores, since increasing core counts beyond a certain point offers diminishing returns.
This configuration is applied during real-time execution.
For new inference engines deployed on already-known devices, we use the configuration that has shown the best performance across the majority of pre-characterized engines. 
For example, when deploying a new engine, we use operating Mode 6 on the AGX node, Mode 9 on the NX node, and 8 threads on the x86 worker.
An Offline characterization phase can later be performed to comprehensively evaluate the new device or inference engine.
The resulting data is then used to enhance the configuration dictionary, improving future scheduling and deployment decisions.

%% file: tables/maths_description.tex
\setlength\extrarowheight{2pt}
\begin{table}[t]
\footnotesize
\caption{Key System and Scheduler Parameters}
\centering
\resizebox{\columnwidth}{!}{%
\begin{tabular}{p{0.15\columnwidth}|p{0.85\columnwidth}}
\specialrule{1.5pt}{0pt}{0pt}
\cellcolor{gray!10}\textbf{Denotation} & \cellcolor{gray!10}\textbf{Short Description} \\
\specialrule{1.5pt}{0pt}{0pt}
$q$ & Number of queries to be executed for the given inference \\
\cline{1-2}
$J$ & Set of inference jobs in the system, where each job $j \in J$. \\
\cline{1-2}
$W$ & Set of worker nodes in the system, where each worker $w \in W$. \\
\cline{1-2}
$c^*_{j, w}$ & Optimal configuration for job $j$ on worker $w$, maximizing QPS. \\
\cline{1-2}
$w^*_{j}$ & The optimal worker for job $j$ that minimizes execution time while meeting QoS constraints. \\
\cline{1-2}
$T_{\text{QoS}, j}$ & Time specified by the user for executing job $j$. \\
\cline{1-2}
$T_{\text{Waiting}, j}$ & Time elapsed since job $j$ was submitted to the queue. \\
\cline{1-2}
$T_{\text{Remaining}, j}$ & Time remaining before a QoS violation occurs for job $j$, calculated as $T_{\text{QoS}, j} - T_{\text{Waiting}, j}$. \\
\cline{1-2}
$T_{\text{Pre-processing}, j}$ & Pre-processing time for job $j$, derived from profiling. \\
\cline{1-2}
$T_{\text{Estimated}, j, w}$ & Estimated execution time for job $j$ on worker $w$, including pre-processing time and execution time per query. \\
\cline{1-2}
$QPS^{c^*_{j, w}}$ & Queries per Second (QPS) achieved by the optimal configuration $c^*_{j, w}$ for job $j$ on worker $w$. \\
\cline{1-2}
$W_{\text{acceptable}, j}$ & Set of workers capable of completing job $j$ within the remaining allowed time. \\
\specialrule{1.5pt}{0pt}{0pt}
\end{tabular}
}
\label{tab:variables}
\end{table}

%% file: sections/04_Experimental.tex
\section{Experimental Evaluation}
\label{sec:evaluation}

In this section, we present the experimental evaluation of \texttt{SynergAI}. 
First, we explain the experimental setup and the mechanisms used for comparison. 
Finally, we present and analyze the results of our experiments and the impact of \texttt{SynergAI}.

\subsection{Experiment Overview and Baselines Description}
\label{ssec:experiments_setup}

We assess the \texttt{SynergAI} scheduler through multiple experiments to evaluate its efficiency.
Each experiment consists of a set of 24 inference engines to be served. 
Each inference engine specifies a required number of queries to be executed along with its QoS constraints, meaning it expects to complete these queries within a specified time limit.
Aiming to evaluate experiments that clearly reflect the behavior of the scheduler, all jobs scheduled and executed in strict isolation on their designated nodes, ensuring zero interference and resource contention.
To determine the demands, we aggregated execution times across all configurations and workers for each inference engine. 
From this distribution, we extracted the median values to define a demand-low-intensity (\texttt{DL}) demand set and the 25\%-ile values to represent a demand-high-intensity workload (\texttt{DH}).
Regarding the arrival of inference requests at \texttt{SynergAI}'s scheduler, we follow a Poisson distribution, as done in prior research~\cite{seo2021slo}, since it effectively models request arrivals by capturing independent events, following an exponential inter-arrival time, reflecting real-world variability, and offering mathematical simplicity for analysis.
The $\lambda$ parameter of the distribution is derived from the data gathered during characterization. 
Specifically, we aggregate execution times for all inference engines across all configurations and workers, extracting the median and 25\%-ile values from the distribution to define a low request frequency (\texttt{FL}) and a high request frequency (\texttt{FH}).
In total, we create three distinct experiments: \texttt{DL-FL}, \texttt{DL-FH}, and \texttt{DH-FH}, each with increasing difficulty to evaluate our scheduling system.

We compare the \texttt{SynergAI} scheduler with various other scheduling methods, ranging from standard approaches to SotA scheduling systems.
Besides our scheduling approach, we implement five additional scheduling systems: i) \textit{Round Robin} (\texttt{RR}), which allocates inference engines to workers in a circular sequence, regardless of worker capabilities or job demands, ensuring a fair distribution of workload across the system; ii) \textit{Strict Round Robin} (\texttt{SRR}), a variation of \texttt{RR}, where each job is strictly assigned to the next worker and waits for its availability, aiming for perfect distribution but risking increased waiting times and demand violations; iii) \textit{Least Recently Used} (\texttt{LRU}), which assigns jobs to the worker that has been idle the longest to prevent starvation and promote an even workload distribution; iv) \textit{Most Recently Used} (\texttt{MRU}), which allocates jobs to the most recently active node that is available, although it risks node starvation, as some nodes may remain underutilized; v) \textit{Best Effort} (\texttt{BE}), which follows a greedy policy, iterating from the strongest worker to the weakest until an available is found, and then assigns the job to that worker for execution. Additionally, we implement from scratch and compare against a SotA scheduling solution derived by~\cite{seo2021slo}. We focus on the proposed solution without model slicing, namely SLO Minimum-Average-Expected-Latency (\texttt{SLO-MAEL}), since our work does not focus on the model slicing techniques. \texttt{SLO-MAEL} aims to reduce QoS violations and the decision-making relies on evaluating all possible mappings of inference engines to workers using a scoring system.
%Specifically, we implemented the \texttt{SLO-MAEL} version as presented, rather than \texttt{PSLO-MAEL}, since our work does not focus on the model slicing techniques that are central to the latter approach.\texttt{SLO-MAEL} seeks to reduce QoS violations of inference engines, similar to \texttt{SynergAI}.It evaluates all possible mappings of inference engines to workers using a scoring system. These scores are calculated based on pre-profiled execution times and the current queue wait times for each worker, which are then used to predict task completion.\texttt{SLO-MAEL} assigns negative scores to mappings that are likely to violate the QoS requirements, prioritizing those that ensure compliance. The mapping with the highest score is then selected for execution.Our implementation of \texttt{SLO-MAEL} adheres to the core principles of \texttt{SLO-MAEL}, with adjustments made to suit our experimental environment. 
The performance of each scheduling system is evaluated using the following metrics: a) total number of violations, b) waiting time, which refers to the duration the inference engine spends in the queue, c) end-to-end execution time, encompassing waiting time, execution time, and the scheduling overhead introduced by \texttt{SynergAI}, and d) average excess time, which measures the average time a job exceeds its desired execution time. 
This is calculated as the difference between the actual and desired times, with excess time clipped to zero for jobs that do not violate the QoS. It is important to note that all jobs are executed isolated, with no concurrent inference workloads on the same node, aiming to avoid interference phenomena.

\iffalse

\subsection{Tuning the Periodic Update Mechanism} Before proceeding to the evaluation, we present the method used to tune the periodic update mechanism, specifically the interval at which queue re-evaluation should be performed. 
To determine this, we executed each of the four available experiments (\texttt{DL/FL}, \texttt{DL/FH}, \texttt{DH/FL}, \texttt{DH/FH}) and analyzed the number of QoS violations for different re-evaluation times. 
We measured this across various re-evaluation intervals of \AF{Add values here...}. 
Figure~\AF{...}

\fi

\subsection{Detailed Analysis of \texttt{SynergAI} versus Defined Baselines}

\begin{figure*}[t]
    \includegraphics[width=0.5\columnwidth]{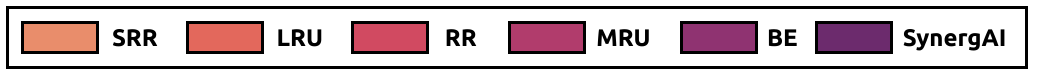}
    \begin{minipage}[t]{\textwidth}
        \captionsetup[subfigure]{labelformat=empty}
        \hspace*{\fill}
        \subfloat[]{
            \includegraphics[width=.24\columnwidth]{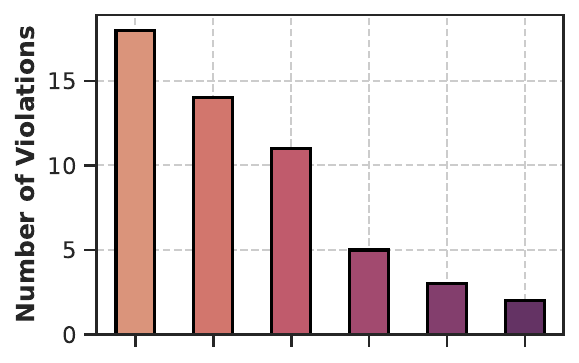}
        }
        \hspace*{\fill}
        \subfloat[]{
            \includegraphics[width=.24\columnwidth]{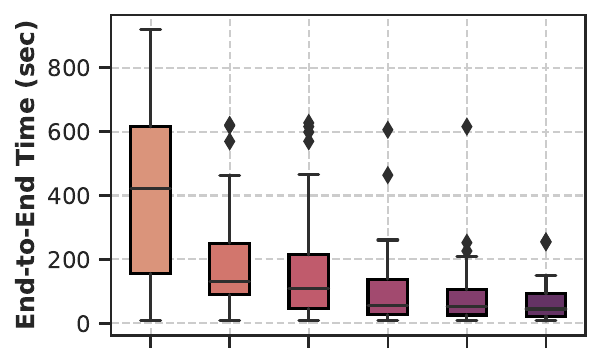}
        }
        \hspace*{\fill}
        \subfloat[]{
            \includegraphics[width=.24\columnwidth]{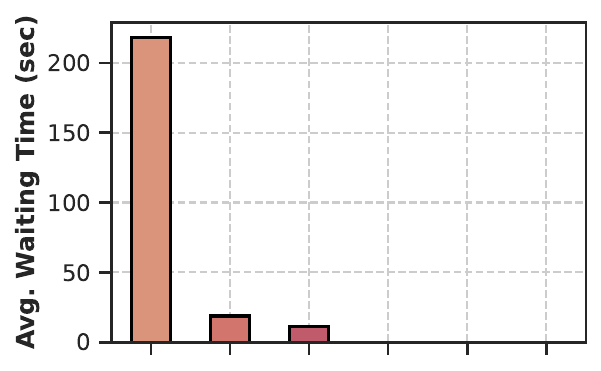}
        }
        \hspace*{\fill}
        \subfloat[]{
            \includegraphics[width=.24\columnwidth]{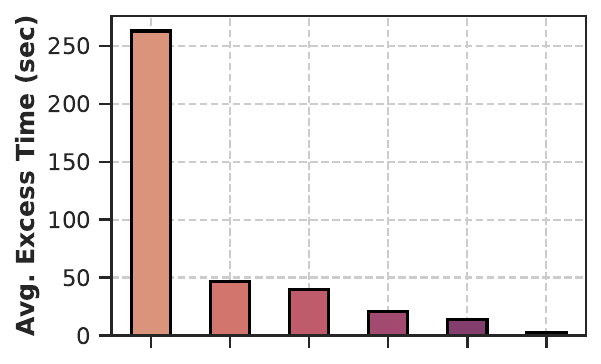}
        }
        \hspace*{\fill}
        \vspace{-20pt}
        \caption{Comparison of \texttt{SynergAI} with other Scheduling Systems in the \texttt{DL/FL} Experiment}
        \label{fig:Evaluation-DL-FL}
    \end{minipage}
\end{figure*}

Figure~\ref{fig:Evaluation-DL-FL} presents the results of the \texttt{DL-FL} experiment.  
Analyzing the number of violations, the \texttt{SRR} scheduler records the highest count at 18, followed by \texttt{LRU} with 14 and \texttt{RR} with 11.  
The \texttt{MRU} scheduler incurs 5 violations, while \texttt{BE} reports 3.  
\texttt{SynergAI} achieves the fewest violations, with only 2 recorded.
Beyond violation counts, \texttt{SynergAI} also demonstrates superior performance in end-to-end execution time.  
Its execution time distribution ranges from a minimum of 7.4 seconds to a maximum of 4.3 minutes, with an average of 1.24 minutes and a tail latency of 4.3 minutes. 
The tail latency metric (i.e. the 99\%-ile of the distribution) is particularly important, as it captures the worst-case execution time for the slowest 1\% of jobs, ensuring system reliability under peak loads.
\texttt{SynergAI} consistently outperforms all other schedulers, with its tail latency execution time being lower by factors of $2.1\times$, $2.3\times$, $2.4\times$, $2.4\times$, and $3.6\times$ compared to \texttt{BE}, \texttt{MRU}, \texttt{RR}, \texttt{LRU}, and \texttt{SRR}, respectively.  
The average waiting time provides further insights into scheduling efficiency.  
As expected, the \texttt{SRR} scheduler exhibits the longest waiting time at 3.6 minutes due to its strict job rotation policy, which evenly distributes workloads but increases delays and violation risks.  
The \texttt{LRU} and \texttt{RR} schedulers show shorter waiting times of 18.8 seconds and 11.5 seconds, respectively.  
Notably, both \texttt{MRU} and \texttt{BE}, along with \texttt{SynergAI}, experience no waiting time.  
\texttt{SynergAI}'s advantage comes from its ability to identify the optimal configuration for each inference engine across worker nodes.  
This underscores the importance of the Offline Phase, as other schedulers rely on predefined configurations, typically selecting the worker with the highest CPU resources.  
By leveraging its learned knowledge, \texttt{SynergAI} surpasses these conventional approaches.  
Moreover, due to its adaptive configuration strategy, even for the two jobs that exceeded their QoS targets, the average excess time remains exceptionally low at just 2.3 seconds.  
This is significantly lower compared to the excess times recorded by other schedulers: 13.4 seconds for \texttt{BE}, 20.7 seconds for \texttt{MRU}, 39.9 seconds for \texttt{RR}, 46.7 seconds for \texttt{LRU}, and 4.3 minutes for \texttt{SRR}.

\begin{figure*}[t]
    \includegraphics[width=0.5\columnwidth]{figures/Evaluation/Evaluation_Legend.pdf}
    \begin{minipage}[t]{\textwidth}
        \captionsetup[subfigure]{labelformat=empty}
        \hspace*{\fill}
        \subfloat[]{
            \includegraphics[width=.24\columnwidth]{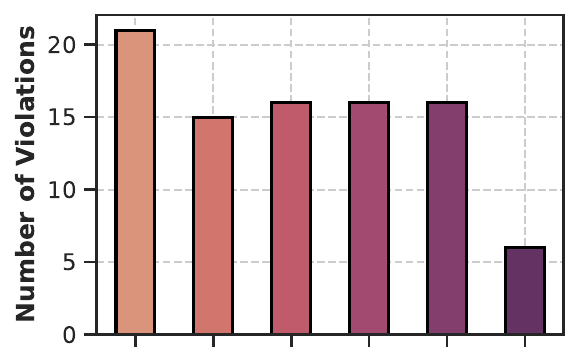}
        }
        \hspace*{\fill}
        \subfloat[]{
            \includegraphics[width=.24\columnwidth]{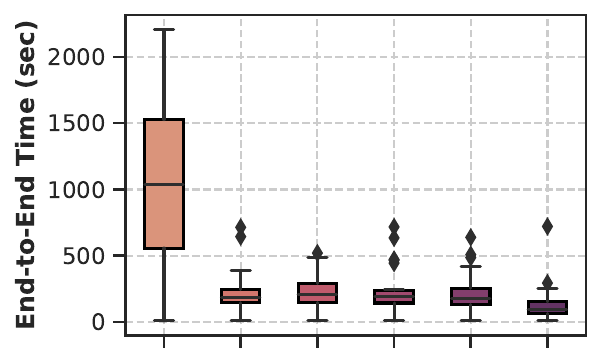}
        }
        \hspace*{\fill}
        \subfloat[]{
            \includegraphics[width=.24\columnwidth]{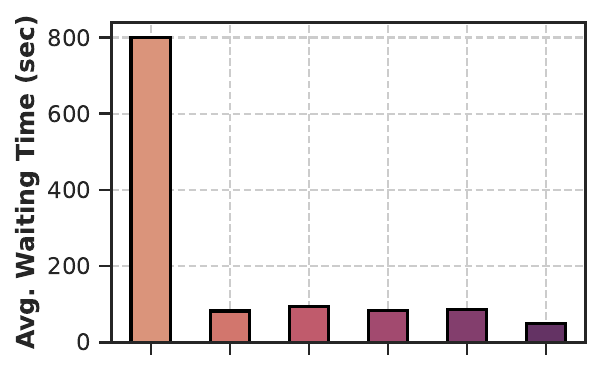}
        }
        \hspace*{\fill}
        \subfloat[]{
            \includegraphics[width=.24\columnwidth]{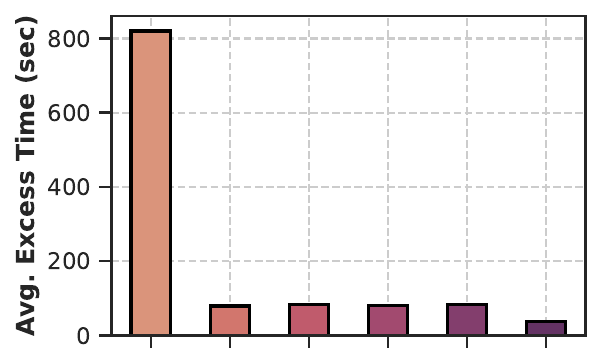}
        }
        \hspace*{\fill}
        \vspace{-20pt}
        \caption{Comparison of \texttt{SynergAI} with other Scheduling Systems in the \texttt{DL/FH} Experiment}
        \label{fig:Evaluation-DL-FH}
    \end{minipage}
\end{figure*}

In the \texttt{DL-FH} experiment, we increase the arrival rate of inference engines to evaluate how the scheduling schemes perform under higher workload conditions.  
Figure~\ref{fig:Evaluation-DL-FH} presents the results of this experiment.  
Among the schedulers, \texttt{SRR} exhibits the highest number of violations, with 21 violations, meaning only three jobs met their constraints.  
The \texttt{LRU} scheduler follows with 15 violations, while \texttt{RR}, \texttt{MRU}, and \texttt{BE} each record 16 violations, indicating a performance decline under heavier load.  
In contrast, \texttt{SynergAI} achieves the best performance, recording the fewest violations at just 6.
For the execution time distribution, \texttt{SynergAI} performs remarkably well, with a range from a minimum of 9.5 seconds to a maximum of 12 minutes, an average of 2.3 minutes, and a tail latency of 10.4 minutes.
Regarding waiting time, \texttt{SRR} has the highest at 13.3 minutes, while \texttt{LRU} and \texttt{RR} exhibit shorter waiting times of 1.36 minutes and 1.56 minutes, respectively.
Both \texttt{MRU} and \texttt{BE} display waiting times of 1.41 minutes and 1.42 minutes, explaining the higher number of violations for these schedulers.  
On the other hand, \texttt{SynergAI} has the smallest average waiting time at just 49.2 seconds.
Finally, regarding excess time, \texttt{SynergAI} maintains an average excess time of 37.7 seconds for the jobs that led to violations. This is significantly lower compared to the excess times recorded by other schedulers: 1.4 minutes for \texttt{BE}, 1.3 minutes for \texttt{MRU}, 1.4 minutes for \texttt{RR}, 1.3 minutes for \texttt{LRU}, and 13.7 minutes for \texttt{SRR}.

\begin{figure*}[t]
    \includegraphics[width=0.5\columnwidth]{figures/Evaluation/Evaluation_Legend.pdf}
    \begin{minipage}[t]{\textwidth}
        \captionsetup[subfigure]{labelformat=empty}
        \hspace*{\fill}
        \subfloat[]{
            \includegraphics[width=.24\columnwidth]{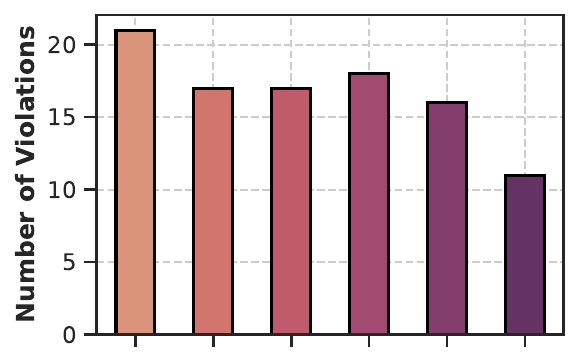}
        }
        \hspace*{\fill}
        \subfloat[]{
            \includegraphics[width=.24\columnwidth]{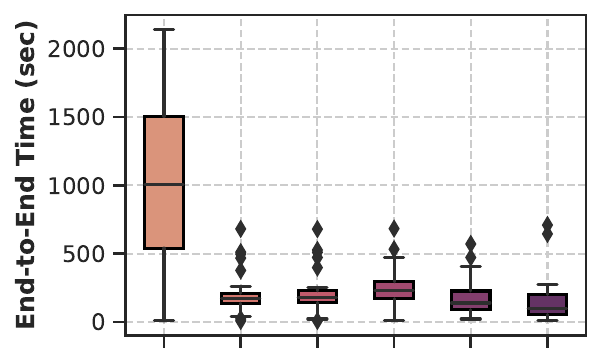}
        }
        \hspace*{\fill}
        \subfloat[]{
            \includegraphics[width=.24\columnwidth]{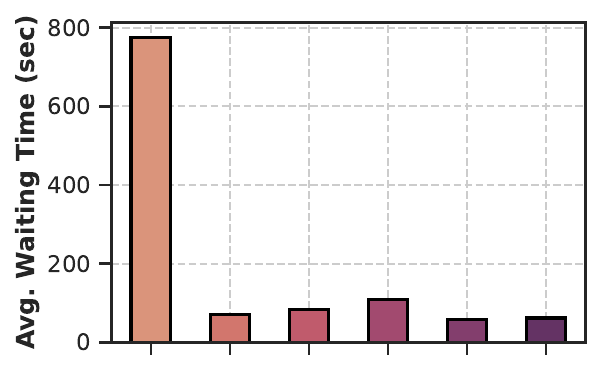}
        }
        \hspace*{\fill}
        \subfloat[]{
            \includegraphics[width=.24\columnwidth]{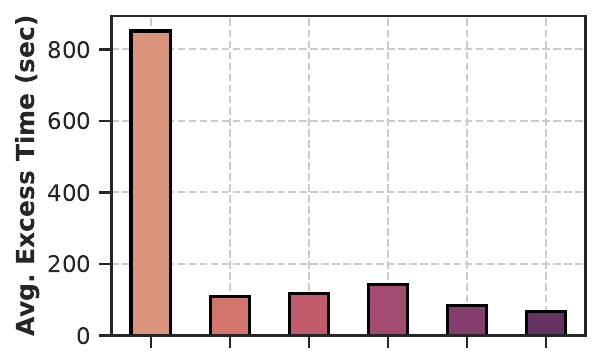}
        }
        \hspace*{\fill}
        \vspace{-20pt}
        \caption{Comparison of \texttt{SynergAI} with other Scheduling Systems in the \texttt{DH/FH} Experiment}
        \label{fig:Evaluation-DH-FH}
    \end{minipage}
\end{figure*}

Finally, in the \texttt{DH-FH} experiment, we not only increase the arrival rate of inference engines but also raise the QoS demands for each engine, making this the most challenging experiment in terms of load for all the schedulers.  
Figure~\ref{fig:Evaluation-DH-FH} presents the results of this experiment.  
Once again, \texttt{SRR} records the highest number of violations, totaling 21.  
The other schedulers show a slight increase in violations compared to the previous experiment, with \texttt{LRU}, \texttt{RR}, \texttt{MRU}, and \texttt{BE} recording 17, 17, 18, and 16 violations, respectively.  
In contrast, \texttt{SynergAI} delivers the best performance even in this demanding scenario, achieving the fewest violations with only 11.  
The execution time distribution of \texttt{SynergAI} ranges from a minimum of 9.3 seconds to a maximum of 11.8 minutes, with an average of 2.75 minutes and a tail latency of 11.6 minutes.  
When comparing the tail latency, \texttt{SynergAI} outperforms the other schedulers, with a value $1.3\times$ lower on average, showcasing its superior efficiency.  
As for waiting times, \texttt{SynergAI} has the lowest average waiting time of approximately 1 minute. 
The next slowest schedulers are \texttt{BE} at 1.1 minutes, followed by \texttt{LRU} at 1.2 minutes, \texttt{RR} at 1.4 minutes, \texttt{MRU} at 1.8 minutes, and \texttt{SRR} at 12.9 minutes.  
Regarding excess time, \texttt{SynergAI} maintains an average excess time of 1.1 minutes for the jobs that resulted in violations, which is significantly lower than the excess times recorded by the other schedulers: 1.4 minutes for \texttt{BE}, 2.35 minutes for \texttt{MRU}, 1.9 minutes for \texttt{RR}, 1.8 minutes for \texttt{LRU}, and 14.2 minutes for \texttt{SRR}.  Summarizing all the experiments, \texttt{SynergAI} achieves an average reduction of $7.1\times$ in QoS violations and $5.3\times$ in excess time, respectively.

%\begin{figure*}[t]
%    \centering
 %   \includegraphics[width=0.5\columnwidth]{figures/Evaluation/Evaluation_Legend.pdf}\\[1ex]
  %  \resizebox{0.4\textwidth}{!}{
   % \includegraphics[width=\columnwidth]{figures/Evaluation/DH-FH/DH_FH_Overhead.pdf}}
   % \caption{Scheduling Overhead of all the Scheduling Systems in the DH/FH Experiment}
   % \label{fig:overhead}
%\end{figure*}

\subsection{Comparison with SotA Scheduling System}

\begin{figure*}[t]
    \includegraphics[width=0.28\columnwidth]{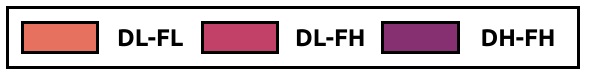}
    \begin{minipage}[t]{\textwidth}
        \captionsetup[subfigure]{labelformat=empty}
        \hspace*{\fill}
        \subfloat[]{
            \includegraphics[width=.24\columnwidth]{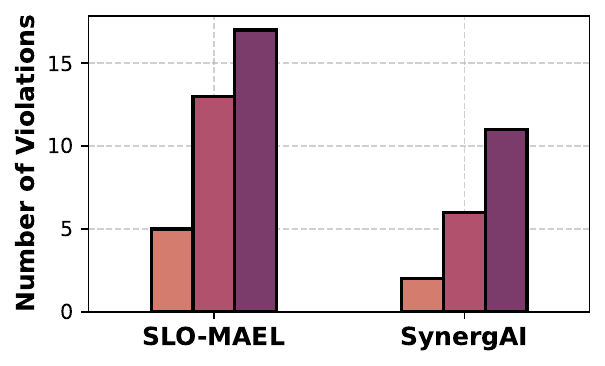}
        }
        \hspace*{\fill}
        \subfloat[]{
            \includegraphics[width=.24\columnwidth]{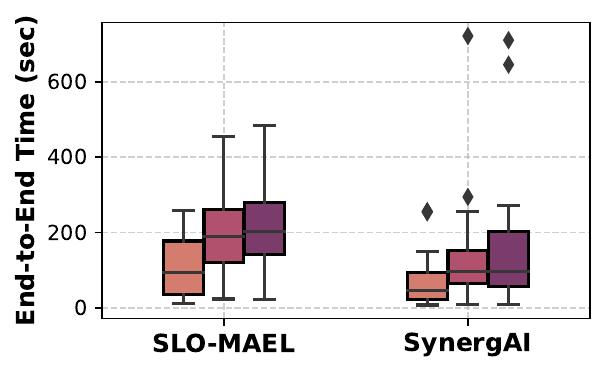}
        }
        \hspace*{\fill}
        \subfloat[]{
            \includegraphics[width=.24\columnwidth]{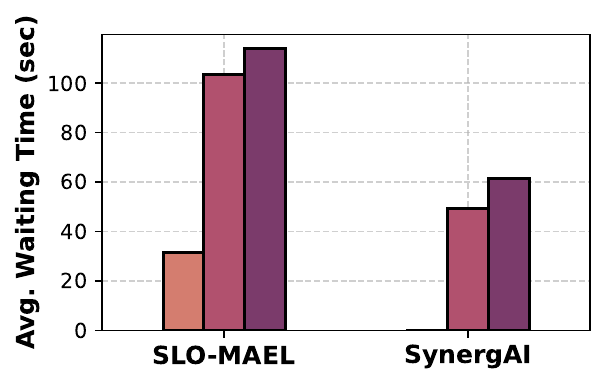}
        }
        \hspace*{\fill}
        \subfloat[]{
            \includegraphics[width=.24\columnwidth]{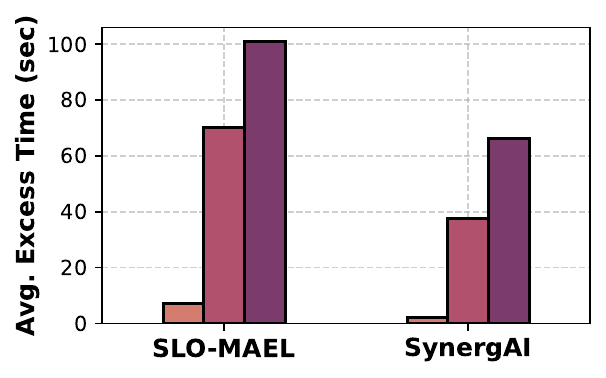}
        }
        \hspace*{\fill}
        \vspace{-20pt}
        \caption{Comparison of \texttt{SynergAI} with SLO Minimum-Average-Expected-Latency (SLO-MAEL)~\cite{seo2021slo} for all Experiments}
        \label{fig:SLO-MAEL}
    \end{minipage}
\end{figure*}

We also compare \texttt{SynergAI} with the SotA scheduling system proposed by~\cite{seo2021slo} (\texttt{SLO-MAEL}). 
We once again compare \texttt{SynergAI} with \texttt{SLO-MAEL} across the three distinct experiments we previously analyzed: \texttt{DL-FL}, \texttt{DL-FH}, and \texttt{DH-FH}, each representing scenarios with escalating difficulty and load. 
For this evaluation, we utilize the same metrics as in our previous analysis to ensure consistency.
Figure~\ref{fig:SLO-MAEL} presents the results from all conducted experiments.  
In the \texttt{DL-FL} experiment, \texttt{SLO-MAEL} results in 5 violations.  
The end-to-end execution time varies significantly, ranging from a minimum of 10.7 seconds to a maximum of 4.4 minutes, with an average of 1.8 minutes and a tail latency of 4.2 minutes.  
In contrast, as previously shown, \texttt{SynergAI} leads to only 2 violations while achieving a $1.5\times$ reduction in average execution time.  
This performance gap can be primarily attributed to the waiting time experienced by tasks.
\texttt{SLO-MAEL} exhibits an average waiting time of 31.5 seconds, whereas \texttt{SynergAI} completely eliminates pending time, allowing for more efficient scheduling and execution.  
Although \texttt{SLO-MAEL} utilizes a scoring system to optimize task-to-worker placement based on current queue conditions, it lacks adaptive rescheduling capabilities and fails to account for the optimal resource configurations of each worker. 
Consequently, this leads to a higher number of violations and longer waiting times.
The drawbacks of \texttt{SLO-MAEL} are also evident in the average excess execution time, which measures how much longer tasks take beyond their expected duration.  
Here, \texttt{SLO-MAEL} records an average excess time of 7.1 seconds, which is $3.1\times$ higher than that of our approach.  
This further highlights the inefficiencies in its scheduling decisions compared to \texttt{SynergAI}, which optimally balances task assignments to minimize delays and violations.

For the \texttt{DL-FH} and \texttt{DH-FH} experiments, which present increasingly complex and challenging scenarios, we observe a consistent pattern.  
\texttt{SLO-MAEL} results in 13 violations for \texttt{DL-FH} and 17 for \texttt{DH-FH}, which are $2.2\times$ and $1.6\times$ higher, respectively, compared to the number of violations seen with \texttt{SynergAI}.  
Execution times are also noticeably higher for \texttt{SLO-MAEL}, with average times that are $1.5\times$ longer for \texttt{DL-FH} and $1.3\times$ longer for \texttt{DH-FH}.  
Similarly, waiting times for \texttt{SLO-MAEL} are significantly greater, standing at $1.9\times$ and $1.5\times$ higher than those observed with \texttt{SynergAI} in these experiments.  
Furthermore, the average excess execution time further emphasizes the inefficiencies of \texttt{SLO-MAEL}, showing $2.2\times$ higher values in \texttt{DL-FH} and $1.85\times$ higher in \texttt{DH-FH} compared to \texttt{SynergAI}.  
These results clearly demonstrate that \texttt{SynergAI} outperforms \texttt{SLO-MAEL} in terms of scheduling efficiency by reducing violations, execution times, waiting times, and excess execution, positioning it as a more effective solution for inference scheduling.

\subsection{Overhead, Energy Consumption Analysis \& Use-Case Demonstration}

\begin{wrapfigure}{r}{6cm}
    \vspace{-7pt}
    \centering
    \includegraphics[width=.4\columnwidth]{figures/Evaluation/Evaluation_Legend.pdf}
    \\
    \includegraphics[width=0.4\columnwidth]{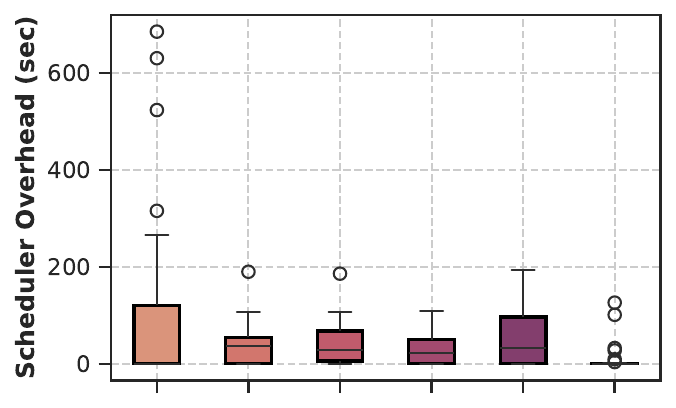}
    \caption{Scheduling Overhead of all the Scheduling Systems in the DH/FH Experiment}
    \label{fig:overhead}
\vspace{-7pt}
\end{wrapfigure}

\textbf{Overhead Analysis}: A successful Edge-Cloud scheduler scheme should also be able to perform decision-making with minimal overhead. 
Therefore, we evaluate the scheduling overhead of each policy in the \texttt{DH-FH} experiment (i.e. the most tight experiment) as shown in Figure~\ref{fig:overhead}, in order to observe the differences in assignment efficiency among the evaluated schedulers. The scheduling overhead represents the time interval from when a job is dequeued from the waiting queue until it is successfully assigned to a worker. 
Note that \texttt{SLO-MAEL} is excluded from the overhead analysis, as its scheduling policy primarily performs decision-making as a pre-processing step for scheduling exploration, thus prior to the actual scheduling decision. Consequently, it does not incur overhead during scheduling, which is the focus of our analysis. Although \texttt{SLO-MAEL} evaluates multiple scheduling scenarios, these computations are not carried out during job execution. Therefore, it is not comparable under our overhead definition, which considers only run-time computational effort. \texttt{SynergAI} demonstrates superior performance with the lowest average scheduling overhead of 12.54 seconds, which is approximately $2.5\times$ faster than MRU, $3.2\times$ faster than LRU, $3.5\times$ faster than RR, $4.1\times$ faster than BE, and $8.9\times$ faster than SRR. The median values further highlight \texttt{SynergAI}'s consistency, with a negligible median scheduling overhead, indicating that the majority of jobs are assigned almost instantaneously.
While SRR also exhibits negligible median scheduling overhead, this metric is misleading and masks its notably poor overall performance. SRR demonstrates an average scheduling overhead of 1.86 minutes -nearly $9\times$ higher than \texttt{SynergAI}- with a maximum scheduling overhead of 11.42 minutes and tail latency of 11.21 minutes, indicating severe efficiency degradation. The deceptive minimal median overhead for SRR can be explained by its strict rotation policy: when the designated worker in the rotation is immediately available, assignment occurs instantly, but when that specific worker is busy, the scheduler waits indefinitely until it becomes available rather than proceeding to the next worker in rotation. This creates a polarized distribution with either immediate assignments or significant delays, making the median an unreliable indicator of SRR's true performance.
\texttt{SynergAI}'s superiority can be attributed to its intelligent multi-layered approach. By computing the best-fitting worker for each job and maintaining a sorted list of workers from fastest to slowest for each specific job, it minimizes assignment overhead through pre-computation and optimization. Additionally, by consistently selecting the fastest available worker for each job, \texttt{SynergAI} ensures rapid job execution, which in turn makes workers available more quickly for subsequent assignments. This creates a positive feedback loop that maintains low scheduling overhead throughout the system's operation. To summarize, \texttt{SynergAI} achieves an average reduction of 4.44× in scheduling overhead across all schedulers.

\begin{figure*}[t]
    \includegraphics[width=0.22\columnwidth]{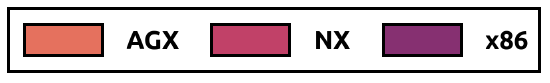}
    \begin{minipage}[t]{\textwidth}
        \captionsetup[subfigure]{labelformat=empty}
        \hspace*{\fill}
        \subfloat[]{
            \includegraphics[width=.48\columnwidth]{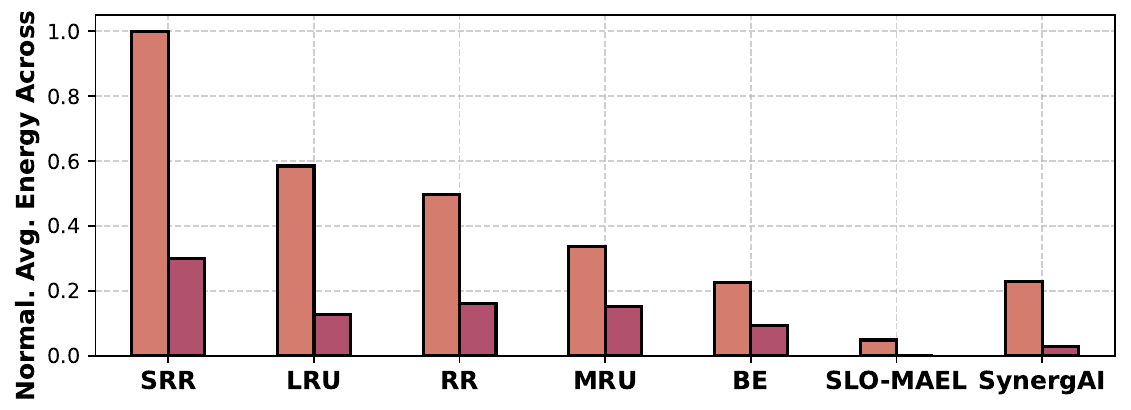}
        }
        \hspace*{\fill}
        \subfloat[]{
            \includegraphics[width=.48\columnwidth]{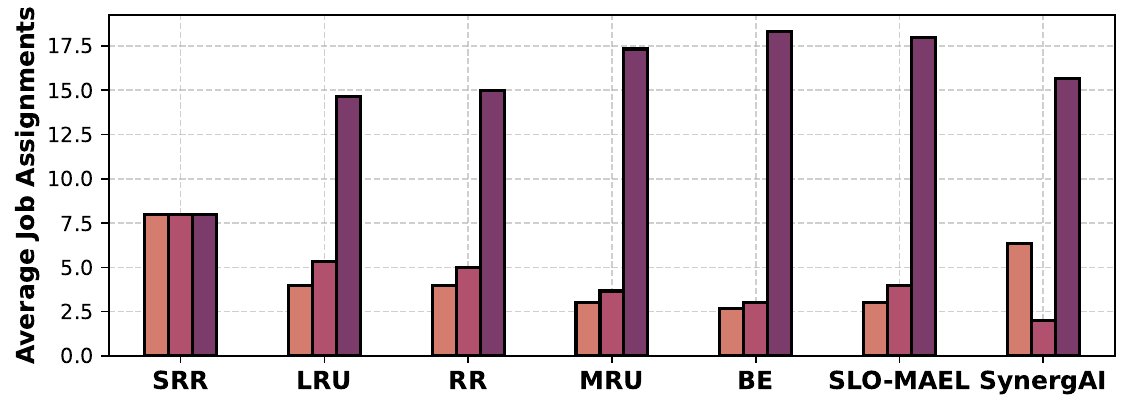}
        }
        \hspace*{\fill}
        \vspace{-20pt}
        \caption{Normalized energy consumption on the Edge nodes (Left) and average job assignment across the Edge-Cloud nodes (Right).}
        \label{fig:energy_all}
    \end{minipage}
\end{figure*}

\textbf{Energy Consumption Analysis}: Figure~\ref{fig:energy_all} presents the normalized average energy consumption on Edge nodes (Left), i.e., the Nvidia AGX and NX boards, and the average job assignment across Edge-Cloud nodes (Right), including the x86 VM. These results cover all the schedulers and experiments discussed in Section~\ref{ssec:experiments_setup}. The Thermal Design Power (TDP) indicates the average power (in Watts) that a processor dissipates when operating at its base frequency with all cores active. Our Intel Xeon server has a TDP of 105\,W per socket, as reported in the manufacturer's datasheet~\cite{xeon}, an order of magnitude higher than the Edge devices, which operate between 10\,W and 30\,W. Consequently, assigning tasks to the x86 worker significantly increases energy consumption. Furthermore, since tasks are deployed on virtual machines (VMs) in the Cloud, accurately attributing energy consumption to a specific VM is not feasible. Thus, this data is omitted from Figure~\ref{fig:energy_all} (left). We observe that \texttt{SynergAI} outperforms the baseline approaches (\texttt{SRR}, \texttt{LRU}, \texttt{RR}, \texttt{MRU}, \texttt{BE}), achieving a 39.08\% reduction in energy consumption on the AGX platform and a 43.42\% reduction on the NX platform, on average. When comparing \texttt{SLO-MAEL} to \texttt{SynergAI}, the former appears to consume less energy, as shown in Figure~\ref{fig:energy_all} (Left). However, Figure~\ref{fig:energy_all} (Right) reveals that \texttt{SLO-MAEL} offloads 14.89\% more jobs on average than \texttt{SynergAI}. Our experiments also demonstrate that jobs offloaded by \texttt{SLO-MAEL} utilize the x86 node for 77.8\% longer than those managed by \texttt{SynergAI}, ultimately leading to higher overall energy consumption across the Edge-Cloud continuum.

\begin{figure}[t]
    \centering
    \includegraphics[width=0.2\columnwidth]{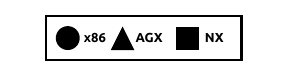}
    \includegraphics[width=\linewidth]{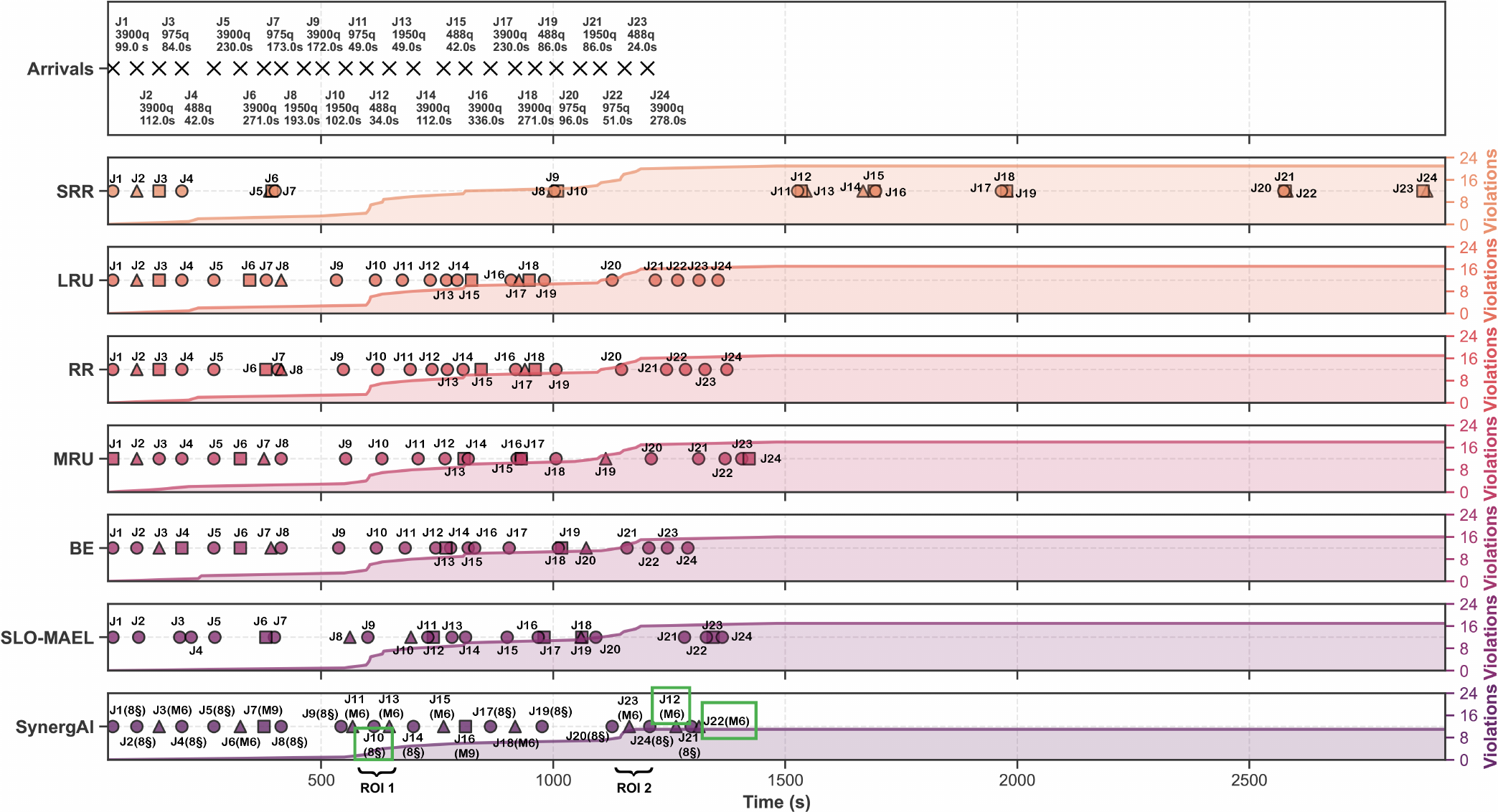}
    \caption{Use Case Illustration for the DH-FH Scenario}
    \label{fig:Use-Case}
\end{figure}

\textbf{Use-case Scenario Breakdown}: We conduct a use case scenario and examine the plots presented in Figure~\ref{fig:Use-Case}, in order to demonstrate how \texttt{SynergAI} performs compared to other scheduling approaches. More specifically, we present a frame-by-frame scenario of our scheduler's behavior under challenging conditions. All the plots are presented over a shared x axis, illustrating the time.
The top plot displays the arrival pattern of inference jobs over time, with each job annotated with its specifications: Job ID ($JX$), number of queries ($q$), and target QoS ($T_{QoS}$). We follow the arrival pattern of the DH-FH experiment, as described in Section~\ref{ssec:experiments_setup}, thus focusing on the most intense experiment with dense arrival regions. The subsequent plots show how different scheduling policies assign the arriving jobs to the available workers over time. Each discrete marker corresponds to the the different nodes examined (i.e. x86, AGX, NX). On the secondary axes, we illustrate the number of violations through time. Each scheduler demonstrates a distinct assignment pattern. The baseline schedulers (\texttt{SRR}, \texttt{LRU}, \texttt{RR}, \texttt{MRU}, \texttt{BE}) assign jobs in their arrival order and jobs are processed sequentially as they arrive, with minimal consideration of job characteristics or system optimization. Similarly, the SoA scheduler, \texttt{SLO-MAEL}, shows some consideration in job placement by evaluating possible mappings, but still maintains a sequential assignment pattern. In addition, these schedulers utilize the default configuration of each device regardless of job requirements. Our analysis focuses into two Regions of Interest (ROIs), i.e. ROI 1 and ROI 2, which represent dense arrival periods, as well as spiking violation regions. \texttt{SynergAI} exhibits fundamentally different behavior through intelligent queue reordering based on job urgency and system state. A clear example is visible with the jobs $J10$, $J11$ and $J12$ in the ROI 1. Job $J12$, arrives early in the sequence, but is strategically delayed and executed much later. We observe also that $J10$, $J11$ are reordered, in contrast to previous solutions. This reordering demonstrates \texttt{SynergAI}'s ability to prioritize jobs based on their urgency rather than simple arrival order, thus leading to reduced QoS violations. Similar observations are derived in ROI 2, with $J21$, $J23$ and $J24$.  Moreover, unlike other schedulers, \texttt{SynergAI} leverages the characterization information to select the optimal configuration for each inference engine on each device, as indicated by the configuration annotations visible in the assignment plot.
Focusing on the secondary axes and the number of violations over time, we observe the critical correlation between arrival patterns and the subsequent increase in total QoS violations across all scheduling methods. The different violation rates among schedulers highlight the importance of intelligent scheduling decisions during high-pressure periods.
\texttt{SynergAI}'s performance in managing these violation surges showcases the effectiveness of its queue reordering and configuration-aware assignment strategy, as it can better distribute the workload and select optimal configurations during critical periods when traditional and SoA schedulers struggle with the increased system pressure. Additionally, this direct correlation between arrival density and violations demonstrates how the Poisson distribution effectively models real-world scenarios where clustered arrivals can overwhelm scheduling systems. Similar behavior has been observed for other use-cases.

%\begin{figure}[t]
 %   \centering
%    \includegraphics[width=0.5\columnwidth]{figures/Use Case/Policy_Legend.pdf}
 %   \includegraphics[width=0.19\columnwidth]{figures/Use Case/Assignment_Legend.pdf}
  %  \includegraphics[width=\textwidth]{figures/Use Case/DH_FH_Timeline_Arrival_SingleLine.pdf}
   % \includegraphics[width=\textwidth]{figures/Use Case/DH_FH_Timeline_Assignment_SRR.pdf}
    %\includegraphics[width=\textwidth]{figures/Use Case/DH_FH_Timeline_Assignment_LRU.pdf}
 %   \includegraphics[width=\textwidth]{figures/Use Case/DH_FH_Timeline_Assignment_RR.pdf}
  %  \includegraphics[width=\textwidth]{figures/Use Case/DH_FH_Timeline_Assignment_MRU.pdf}
   % \includegraphics[width=\textwidth]{figures/Use Case/DH_FH_Timeline_Assignment_BE.pdf}
    %\includegraphics[width=\textwidth]{figures/Use Case/DH_FH_Timeline_Assignment_SLO-MAEL.pdf}
 %   \includegraphics[width=\textwidth]{figures/Use Case/DH_FH_Timeline_Assignment_SynergAI.pdf}
  %  \includegraphics[width=\textwidth]{figures/Use Case/DH_FH_Timeline_Violation_Lineplot.pdf}
  %  \vspace{-20pt}
  %  \caption{Use Case Scenario}
  %  \label{fig:Use-Case}
%\end{figure}

%\MK{If we have the metrics, as SX proposed we could provide an analysis on the accuracy of the mathematical model, i.e. showing how much we estimated for an engine to run and how much it was actually executed.}

%% file: sections/05_Conclusion.tex
\section{Conclusion \& Future Work}
\label{sec:conclusion}

In this work, we introduce \texttt{SynergAI}, an intelligent scheduling framework designed to dynamically optimize workload distribution across heterogeneous edge and cloud environments.  
By leveraging extensive performance characterization, \texttt{SynergAI} efficiently allocates inference-serving workloads to minimize QoS violations while maximizing resource utilization.  
Our framework accounts for the trade-offs between performance and architecture-operating modes, ensuring a well-balanced deployment strategy across the computing continuum.
Through seamless integration within a Kubernetes-based ecosystem, \texttt{SynergAI} demonstrates its effectiveness in handling diverse inference-serving scenarios, adapting to varying workloads, and improving overall system efficiency.  
Our findings show that architecture-driven inference serving facilitates optimized, efficient deployments on emerging hardware platforms, resulting in an average reduction of $2.4\times$ in QoS violations compared to a SotA solution.

For future work, we plan to extend \texttt{SynergAI} with GPU-enabled workloads across the Edge-Cloud continuum. Going beyond CPU-driven inference, there exist scenarios where real-time inference is critical, thus accelerating through GPU enables an additional level of exploration. Beyond conventional DNNs, our objective is to support increasingly complex models, including Large Language Models (LLMs) and multi-modal architectures, which require significantly higher computational throughput. 
To facilitate this, we also plan to integrate automated DNN partitioning capabilities, enabling \texttt{SynergAI} to dynamically split and distribute neural network components among heterogeneous Edge devices based on their current load and hardware characteristics. 
Furthermore, we will investigate the impact of resource interference, arising from co-located workloads, on inference performance and scheduling efficiency. 
This combined approach, leveraging GPU acceleration, intelligent model partitioning, and interference-aware scheduling, will further enhance scalability, reduce latency and energy consumption, and minimize QoS violations across diverse deployment scenarios.